\magnification=\magstep1          \overfullrule=0pt
\advance\hoffset by -0.35truecm
\font\tenmsb=msbm10    \font\sevenmsb=msbm7  \font\fivemsb=msbm5    
\newfam\msbfam   \textfont\msbfam=\tenmsb  \scriptfont\msbfam=\sevenmsb
\scriptscriptfont\msbfam=\fivemsb     \def\Bbb#1{{\fam\msbfam\relax#1}}
\def\Q{{\Bbb Q}}\def\R{{\Bbb R}}\def\Z{{\Bbb Z}}
\def\C{{\Bbb C}}           
\font\klein=cmr8     \font\itk=cmti8       \font\bfk=cmbx8          
\font\gross=cmr10 scaled \magstep2
\font\kap=cmcsc10  
\newread\epsffilein    \newif\ifepsffileok    \newif\ifepsfbbfound   
\newif\ifepsfverbose   \newdimen\epsfxsize    \newdimen\epsfysize    
\newdimen\epsftsize    \newdimen\epsfrsize    \newdimen\epsftmp      
\newdimen\pspoints  \pspoints=1bp  \epsfxsize=0pt  \epsfysize=0pt         
\def\epsfbox#1{\global\def\epsfllx{72}\global\def\epsflly{72}%
 \global\def\epsfurx{540}\global\def\epsfury{720}%
 \def\lbracket{[}\def\testit{#1}\ifx\testit\lbracket
 \let\next=\epsfgetlitbb\else\let\next=\epsfnormal\fi\next{#1}}%
\def\epsfgetlitbb#1#2 #3 #4 #5]#6{\epsfgrab #2 #3 #4 #5 .\\%
 \epsfsetgraph{#6}}%
\def\epsfnormal#1{\epsfgetbb{#1}\epsfsetgraph{#1}}%
\def\epsfgetbb#1{\openin\epsffilein=#1
\ifeof\epsffilein\errmessage{I couldn't open #1, will ignore it}\else
 {\epsffileoktrue \chardef\other=12
  \def\do##1{\catcode`##1=\other}\dospecials \catcode`\ =10 \loop
  \read\epsffilein to \epsffileline \ifeof\epsffilein\epsffileokfalse\else
       \expandafter\epsfaux\epsffileline:. \\    \fi
  \ifepsffileok\repeat   \ifepsfbbfound\else
  \ifepsfverbose\message{No bounding box comment in #1; using defaults}\fi\fi
  }\closein\epsffilein\fi}%
\def\epsfclipstring{}%

\def\epsfsetgraph#1{ \epsfrsize=\epsfury\pspoints 
 \advance\epsfrsize by-\epsflly\pspoints  \epsftsize=\epsfurx\pspoints
 \advance\epsftsize by-\epsfllx\pspoints \epsfxsize\epsfsize\epsftsize
  \epsfrsize
\ifnum\epsfxsize=0\ifnum\epsfysize=0\epsfxsize=\epsftsize\epsfysize=\epsfrsize
   \epsfrsize=0pt \else\epsftmp=\epsftsize \divide\epsftmp\epsfrsize
  \epsfxsize=\epsfysize \multiply\epsfxsize\epsftmp
  \multiply\epsftmp\epsfrsize \advance\epsftsize-\epsftmp \epsftmp=\epsfysize
  \loop \advance\epsftsize\epsftsize \divide\epsftmp 2 \ifnum\epsftmp>0
  \ifnum\epsftsize<\epsfrsize\else \advance\epsftsize-\epsfrsize
  \advance\epsfxsize\epsftmp \fi \repeat \epsfrsize=0pt \fi \else
  \ifnum\epsfysize=0 \epsftmp=\epsfrsize \divide\epsftmp\epsftsize
  \epsfysize=\epsfxsize \multiply\epsfysize\epsftmp
  \multiply\epsftmp\epsftsize \advance\epsfrsize-\epsftmp \epsftmp=\epsfxsize
  \loop \advance\epsfrsize\epsfrsize \divide\epsftmp 2 \ifnum\epsftmp>0
  \ifnum\epsfrsize<\epsftsize\else \advance\epsfrsize-\epsftsize
  \advance\epsfysize\epsftmp \fi \repeat \epsfrsize=0pt \else
  \epsfrsize=\epsfysize \fi \fi \ifepsfverbose\message{#1:
  width=\the\epsfxsize, height=\the\epsfysize}\fi \epsftmp=10\epsfxsize
  \divide\epsftmp\pspoints \vbox to\epsfysize{\vfil\hbox to\epsfxsize{
  \ifnum\epsfrsize=0\relax \includegraphics{#1} \else \epsfrsize=10\epsfysize \divide\epsfrsize\pspoints
  \includegraphics{#1}\fi \hfil}}%
\global\epsfxsize=0pt\global\epsfysize=0pt}%
{\catcode`\%=12 \global\let\epsfpercent=
\long\def\epsfaux#1#2:#3\\{\ifx#1\epsfpercent
   \def\testit{#2}\ifx\testit\epsfbblit    \epsfgrab #3 . . . \\%
      \epsffileokfalse     \global\epsfbbfoundtrue
   \fi\else\ifx#1\par\else\epsffileokfalse\fi\fi}%
\def\epsfempty{}\def\epsfgrab #1 #2 #3 #4 #5\\{%
\global\def\epsfllx{#1}\ifx\epsfllx\epsfempty \epsfgrab #2 #3 #4 #5 .\\\else
   \global\def\epsflly{#2} \global\def\epsfurx{#3}\global\def\epsfury{#4}\fi}
\def\epsfsize#1#2{\epsfxsize} 
\def\qed{{\vrule height4pt width4pt depth1pt}}\def\hfbr{\hfill\break} 
\def\oddots{\mathinner{\mkern1mu\raise1pt\vbox{\kern7pt\hbox{.}}
  \mkern2mu\raise4pt\hbox{.}\mkern2mu\raise7pt\hbox{.}\mkern1mu}}
\def\lb{\lbrack}\def\rb{\rbrack}  \def\q#1{$\lb {\rm #1}\rb$}
\def\bn{\bigskip\noindent} \def\mn{\medskip\smallskip\noindent}
\def\sn{\smallskip\noindent} \def\one{{\bf 1}} \def\n{^{(n)}} 
\def\a{\hbox{$\cal A$}}  \def\w{\hbox{$\cal W$}}
\def\h{\hbox{$\cal H$}}  \def\p{\hbox{$\cal P$}}
\def\g{\hbox{$\cal G$}}  \def\b{\hbox{$\cal B$}}
\def\f{\hbox{$\cal F$}}  \def\V{\hbox{$\cal V$}} 
\def\lra{\longrightarrow} 
\def\us{^{{\rm s}}} \def\ua{^{{\rm a}}} \def\uh{^{{\rm h}}}
\def\hwr{highest weight representation}
\def\kapp{{\kap Proof}: }    \def\ctq{\hbox{$c(2,q)$}}  
\def\bfctq{\hbox{$c(2,q)$}\kern-26.1pt\hbox{$c(2,q)$}}  
         \def\nvs{\noalign{\vskip 5pt}}
{\nopagenumbers
\line{ETH-TH/96-44 \hfill hep-th/9612178}
\bn\bn\bn\bn
\centerline{{\gross {}From Path Representations to Global Morphisms}} 
\bn
\centerline{{\gross for a Class of Minimal Models}} 
\bn\sn
\vfill \centerline{by}\bn\mn
\centerline{{Andreas Recknagel}}
\bn\sn
\centerline{
Theoretische Physik, 
ETH-H\"onggerberg}
\centerline{CH-8093 Z\"urich,  
Switzerland}
\bn\bn\vfill
\bn\bn\bn
\centerline{{\bf Abstract}}
\bn
\narrower
We construct global observable algebras and global DHR morphisms for the 
Virasoro minimal models with central charge \ctq, $q$ odd. To this end, we 
pass {}from the irreducible highest weight modules to path representations,  
which involve fusion graphs of the \ctq\ models. The paths have an 
interpretation in terms of quasi-particles which capture some structure 
of non-conformal perturbations of the \ctq\ models. The path algebras 
associated to the path spaces serve as algebras of bounded observables. 
Global morphisms which implement the superselection sectors are constructed 
using quantum symmetries: We argue that there is a canonical semi-simple
quantum symmetry algebra for each quasi-rational CFT, in particular for the  
\ctq\ models. These symmetry algebras act naturally on the path spaces, which 
allows to define a global field algebra and covariant multiplets therein. 
\bn
\bn\bn
\vfill\vfill\vfill
\sn
e-mail address: anderl@itp.phys.ethz.ch  
\eject
\vfill $$\phantom{Leerseite}$$\vfill
\eject} 
\pageno=1  
\leftline{\bf 1. Introduction}
\bn
The algebraic approach to quantum field theory as developed by Haag, Kastler,
Doplicher and Roberts \q{37,18} aims at a formulation of quantum theory in
terms of observable quantities only: A theory is defined by a net of local
observables, i.e.\ an assignment $I \lra
\a(I)$, where $\a(I)$ is the ${}^*$-algebra generated by all 
bounded ``measurements'' (observables) that can be made in some 
bounded region $I$ of space-time. This net is subject to certain
explicit axioms  like isotony and causality (or rather Haag duality). 
Different superselection sectors of the quantum theory correspond 
to inequivalent positive energy representations of the local net 
on Hilbert spaces $\h_i$; the representations have to be covariant 
with respect to some group of space-time transformations (including 
the Poincar\'e group); among the representations, there is a 
vacuum sector $\h_0$. With the help of a localization criterion --  
which a priori rules out charges of gauge theories, but is 
tailor-made for short-range strong interactions as well as for 
charges occurring in $1+1$ dimensional conformal field theories -- 
Doplicher, Haag and Roberts were able to determine a class of 
representations $\pi_i$ which can be described by algebra 
endomorphisms $\rho_i$ of the local net: $\pi_i \simeq \pi_0\circ
\rho_i\,$. {}From the observable algebra and the endomorphisms, 
one can reconstruct a field algebra containing charged operators. \hfbr
\noindent The DHR formulation offers a good framework for abstract and  
mathematically precise investigations. With the endomorphisms, it 
provides a unique and efficient tool absent {}from any other approach to 
QFT. In particular, the DHR endomorphisms allow for a natural definition 
of ``fusion'' of representations simply as composition of 
endomorphisms, and they make it possible to study the statistical 
properties of superselection sectors in a quantitative fashion, 
culminating in the duality theorem of 
Doplicher and Roberts which completely classifies  the 
fusion structures that are realized by local QFT 
in $3+1$ dimensions \q{19}. For precise statements and 
further details, see e.g.\ \q{36} and the references quoted there. \hfbr
\noindent Although  originally designed for the study of QFT 
in physical Minkowski space-time, the DHR framework can be applied 
to conformal field theory in $1+1$ dimensions as well, without the need 
to change the basic axioms \q{24}. Fredenhagen, Rehren and Schroer could 
show that  algebraic QFT provides an entirely natural 
and computationally effective treatment of the phenomenon of braid 
group statistics \q{27} typical for low-dimensional quantum 
field theories, and again the DHR endomorphisms prove very effective 
for classifying the braid group representations realized 
by the sectors, see also \q{28,52}. \hfbr
\noindent The DHR approach has led to a number of theorems on the 
general structure of chiral conformal field theories; here we mention only
Longo's construction of Jones inclusions {}from endomorphisms
\q{43,55} and Wiesbrock's results \q{64} -- based on earlier work by 
Borchers and others -- on the realization of the  M\"obius group 
SU(1,1) through the Tomita-Takesaki modular operators, see also \q{33}. 
\sn
On the other hand, the ``conventional approach'' to CFT, initiated 
in the work \q{8}, and in \q{26,42,65}, has its merits, too. Compared 
to the DHR formulation, it is conceptually less stringent and less 
rigorous mathematically, but much more successful when it comes to the 
description of specific models. A theory is defined in terms of 
explicit generators (unbounded Laurent modes of point fields) and 
(commutation) relations, and representation theory of Lie algebras 
plays a prominent role. In contrast to the von Neumann algebra 
approach of the algebraic formulation, unitary and non-unitary 
models can essentially be treated on the same footing. The study 
of models has led to a variety of interesting results which have no 
counterpart in the algebraic approach, among them the construction 
of $W$-algebras and the discovery of the role of the modular group, 
which poses severe restrictions onto the possible space of states. 
Beyond that, it has been possible to fit CFT into the wider frame 
of string theory and of two-dimensional integrable models, including 
models of statistical mechanics on the lattice. \hfbr
\noindent If one had been limited to the investigation of general 
properties without the possibility to analyze specific models, 
many of these aspects would probably have gone unnoticed. Therefore, 
we think that it would be beneficial for algebraic QFT to try to 
discuss models in its language;  in turn, the conceptual power of 
the DHR approach should  provide valuable new insights e.g.\ for 
the complete classification of conformal field theories. 
\sn
Model-building is rather problematical in the algebraic framework. 
Starting {}from scratch with an abstract net of von Neumann algebras 
and identifying the  corresponding conformal model only {}from the 
relative positions of the local algebras, has, to our knowledge, 
never been carried out in concrete examples. Alternatively, one can 
take a model given in the ``conventional'' formulation and try to 
``translate'' into the algebraic language. The ``dictionary'', 
see e.g.\ \q{30}, is easily set up for some parts of the theory, e.g.\ 
the statistical structure, and even the construction of a local net 
is, formally, straightforward: One smears out the point fields with 
test functions supported in an interval $I\subset S^1$ and forms 
bounded functions of the resulting operators; these generate $\a(I)$.  
Covariance of the net is guaranteed if one starts with unbounded 
fields that are (quasi-)primary with respect to the central 
extension of Diff$(S^1)$. However, in forming such local algebras, 
the advantages of working with concrete (unbounded) generators and 
relations are lost, and it is in general not possible to construct 
DHR morphisms explicitly. 
\sn
A few conformal models can be treated in the 
DHR framework, at least to some extent. The simplest example is the 
``free boson'', more precisely the CFT associated to the U(1) current 
algebra, which is discussed in great detail in \q{15}. There, the 
local algebras $\a(I)$ have (bounded) generators obeying Weyl type 
relations, which are simple enough to gain complete control of 
the local net, and one can also explicitly determine the localized 
DHR endomorphisms (which are automorphisms in this model). 
In the cases of non-abelian WZW models, loop group representation 
theory is a useful tool for the study of local nets \q{33,63}, 
but DHR endomorphisms have not been constructed along these lines. 
\sn
A slightly different approach has been used in \q{44}, where the 
Ising model was reformulated by algebraic methods. One can again 
introduce local observable algebras as above, but for the explicit 
construction of DHR endomorphisms 
Mack and Schomerus exploited the fact that there is a natural 
$C^*$-algebra associated to the Ising model: The maximally extended 
chiral symmetry algebra of the Ising model is generated by a free 
fermion, the Laurent modes $b_r$ of which are bounded operators and 
form a global algebra of bounded observables. More precisely, 
observables are bilinear in the fermion modes, and the full algebra 
decomposes into the direct sum of Neveu-Schwarz and Ramond sector. 
Since the generators $b_r$ satisfy simple anti-commutation relations, 
it is not too difficult to give formulas for 
global endomorphisms. In \q{44}, it was shown that these implement 
the superselection sectors and fusion rules familiar {}from the 
Ising model; furthermore, a field algebra with quantum symmetry 
action was constructed and the braid group representations associated 
with the superselection structure were identified. The authors also 
gave plausible arguments why their global endomorphisms are equivalent 
to localized ones, which was proven only later in \q{13} 
using  more elaborate operator algebraic techniques.
\sn
The results of \q{44} were useful for a number of other models, 
too, since there exist free fermion representations of 
e.g.\ all level 1 WZW models \q{31}. The morphisms of these 
theories are very similar to those of the Ising model. With 
additional effort, using ``multi-flavor'' fermion representations, 
one can also describe some aspects of certain higher level WZW 
models in algebraic terms \q{14}. 
\mn
In the case of a conformal QFT, the von Neumann algebras forming 
a local net are all isomorphic to the hyperfinite factor of type
III${}_1$ and display no ``individual structure''. The global 
$C^*$-algebra of \q{44}, on the other hand, does have some
``model-characterizing structure'' which, to some extent,
substitutes for the explicit generators and relations of the 
conventional approach. If we want to find an algebraic 
description of other conformal theories like the 
minimal models of the Virasoro algebra, and in particular want to 
find concrete formulas for DHR morphisms, we think that it is most 
promising to follow this $C^*$-algebraic approach. However, in trying 
to do so, we meet difficulties at once, simply because, unlike the 
Ising model, the chiral algebras of the other Virasoro minimal models 
have no generating fields with bounded Laurent modes. Thus, we have 
to look for a new device to supply us with a $C^*$-algebra of 
bounded observables which can be used as a starting point. 
\sn
Such a device is available for the special sub-series of minimal 
models with central charge $c(2,q)$, $q$ odd: Instead of working 
with Virasoro modes directly, we first introduce {\sl path 
representations} for the irreducible highest weight modules of the 
\ctq\ models. These are vector spaces generated by (finite) paths 
on certain graphs. The graphs carry labels which make it possible to
define an ``energy grading'' on the path spaces, and one can show that for
each irreducible \ctq\ module there is a labeled graph such that the spectrum
of the path energy operator coincides with the true energy spectrum in the 
CFT highest weight representation. Thus, the most important, and the only
invariant, information contained in the state spaces can be captured on a
labeled graph. It turns out that one may use the same graph for all sectors 
of a given model and that only the path starting conditions depend on the
sector. Incidentally, the relevant graph is the fusion graph of a certain
primary field in the given \ctq\ model.
\sn
There is evidence that the path representations themselves 
are physically relevant: They admit an interpretation in terms 
of quasi-particles which contains information on non-conformal 
deformations of the \ctq\ models. But we can also view the construction    
of the path spaces as a mere representation theoretic interlude 
(of a somewhat unusual kind), and exploit it in order to obtain a 
natural global $C^*$-algebras of bounded observables that characterize 
our models: There is a canonical $C^*$-algebra acting on each path space, 
namely the {\sl path algebra} over the respective graph. We take this 
algebra as the global observable algebra \a\ in the corresponding 
representation, and identify the universal global observable algebra with 
the direct sum of these algebras over all sectors. As in the Ising case, 
\a\ has non-trivial center, in agreement with the general theorems of 
\q{24}: For chiral conformal QFTs, the center of \a\ contains global 
charges (one per sector), and the vacuum representation is not faithful. 
Basically, this happens because on the ``space-time'' $S^1$ one cannot 
``hide an electron behind the moon''. 
\sn
There is another structural similarity between our \a\ and the global 
Ising observable algebra, a more surprising one in view of the very 
different constructions: In both cases, the global algebras are 
{\sl AF-algebras}, i.e.\ inductive limits of finite-dimensional 
matrix algebras. This property is obvious for the path algebras of 
the \ctq\ models, whereas in the Ising model it can be seen after 
expressing bi-linears in the fermion modes through 
Temperley-Lieb-Jones generators, see \q{44}. \hfbr
\noindent Moreover, the algebras associated to different sectors 
of a given model -- Ising or one of the \ctq\ series -- are mutually
``stably isomorphic'', i.e.\ they are isomorphic after tensoring with 
the compact operators. Thus, it appears that the stable isomorphism 
type of the global algebras characterizes the conformal model.
\mn
We believe that this is in fact a general phenomenon, and that 
to each rational conformal model there is an AF-algebra playing 
the role of the global observable algebra. Although we have no 
proof for this conjecture, there is some amount of evidence: 
Already within the algebraic approach, one is naturally led to 
study the AF-sub-algebras of the observable algebra which are 
generated by intertwiners \q{24,28}. These contain all information 
on the statistics of the model, and in view of the very restrictive
conformal covariance laws they determine at least a great 
deal of the global observable algebra of a CFT. Other examples of 
AF-algebras are provided by various kinds of lattice approximations 
of conformal models. The two-dimensional RSOS lattice models, see
e.g.\ \q{6,50}, are by definition based on path spaces (in a sense 
different {}from ours). 
Similarly, the lattice current algebras introduced by Faddeev, which are 
one-dimensional discretizations of conformal models, display an 
AF-structure \q{2}, which is somewhat more subtle but should become 
important precisely for studying the continuum limit of these 
lattice models. Since one expects to ``obtain'' conformal field theories 
in this limit, both {}from RSOS models and {}from lattice current algebras,   
it is natural to conjecture that AF-algebras show up in the 
continuum theory as well. One might hope that eventually conformal 
field theories can be classified by {\sl invariants} of ``their associated'' 
AF-algebras (plus some extra data) similar to the classification of 
Lie algebras by Dynkin diagrams. 
\mn
Coming back to the comparison of the Ising with the \ctq\ models, 
we observe that in our cases it is much more difficult to define 
local sub-algebras of the global observable algebra \a\ explicitly. 
In the Ising model, the fermion modes can be used both to generate the 
global observable algebra and to form a covariant local net. In our 
cases, one could build local algebras {}from the (smeared) energy 
momentum field, but these will be sub-algebras of the global path 
observable algebra only if we start {}from a Virasoro action on the 
paths. As we will see, the construction of the path representations 
``imports'' a seemingly natural operation of the 
Virasoro algebra into the path spaces, but this action is 
very irregular in terms of paths and virtually impossible 
to control. One need not stick to this ``natural'' Virasoro action 
and can instead look for another implementation of Diff$(S^1)$ or 
SU(1,1) which takes the specific structure of the path spaces into 
account and is, accordingly, easier to handle within the path algebra. 
Progress towards such representations of SU(1,1) 
has been made in \q{56}, but the results are not yet explicit 
enough to construct covariant local nets. 
In view of this, we have to be content with an algebraic treatment 
of the global aspects of the \ctq\ minimal models at present.
\mn
The next aim is to construct (global) {\sl morphisms} of the (global) 
path observable algebra which implement the superselection sectors. 
Again, we cannot simply copy the procedure of \q{44} used for the 
Ising model, where it was the existence of the free fermion modes 
which allowed to ``guess'' the right endomorphisms.  In the 
work \q{62}, however, Vecserneys gave an alternative construction of 
morphisms for the Ising model, starting {}from an action of 
a semi-simple quantum symmetry algebra (QSA) on the Hilbert space of 
states. In this context, the morphisms necessarily arise in the form of 
(non-unital) {\sl amplimorphisms} of the observable algebra -- i.e.\ as 
maps {}from \a\ into some matrix amplification $M_n(\a)$ -- and not as 
(unital) endomorphisms. The representations they implement 
are, however, equivalent to the ones obtained in \q{44}. 
\sn
Here, we will follow the procedure of \q{62} and use an action of a 
suitably chosen quantum symmetry algebra on the path 
spaces associated to a \ctq\ model. For our purposes, we will not need the 
full structure of a QSA but only the information encoded in a co-product. 
The action of the QSA on the \ctq\ path spaces leads to a natural global 
path field algebra \f\ and to equations defining covariant field 
multiplets, which are elements in amplifications of \f. The charged 
multiplets immediately yield the desired amplimorphisms. 
\bn
The paper is organized as follows: In the next section, we list the 
basic data of the \ctq\ minimal models and then show how to construct 
path representations of the irreducible modules and of the global 
observable algebra. In addition, we present a physical interpretation
of the path spaces in terms of quasi-particles. Section 3 starts with a 
few general remarks on quantum symmetries of low-dimensional QFTs; returning 
to our models, we make a special choice for the quantum symmetry algebra 
for the \ctq\ theories and implement it on the path spaces -- which leads 
to a natural definition of the global field algebra. In section 4, we 
first construct amplimorphisms of the QSA, which are used to formulate 
equations for covariant multiplets within the field algebra; finding 
solutions to these equations, which is achieved by a careful analysis of 
the combinatorial structure of the path spaces, is the main subject of 
that section. The then straightforward definition of global amplimorphisms 
of the path observable algebra is given in section 5, followed by a short 
discussion of properties of these maps. We conclude with an outlook 
on future developments. Subsections 2.3 on quasi-particles and 
also 3.2, where we justify the special (canonical) choice of our QSA, 
can be regarded as digressions which are not essential for the 
construction of amplimorphisms, but in our view they contain some 
considerations which are interesting in their own rights.
\bn\bn
\vfill\eject
\leftline{\bf 2. Path representations} 
\bn
We first set up notations and collect some facts about the \ctq\ minimal
models which will be used as input data for the following constructions.
In subsection 2.2, we start with a brief review of some results of 
\q{22}, where explicit bases for the irreducible modules of the \ctq\ 
models were obtained. {}From there, it is relatively easy to give a 
re-interpretation of the FNO bases in terms of paths on certain 
graphs \q{40}. The resulting path representations probably bear a deeper 
physical meaning, which will be commented on in section 2.3. The 
immediate consequence relevant for our purposes is that they provide 
a partial operator algebraic description of the \ctq\ conformal models 
which makes at least some of the DHR concepts applicable. 
\bn\bn
{\bf 2.1 Data of the \bfctq\ minimal models}
\bn
The conformal models we are going to study possess a maximal chiral 
observable algebra (the ``energy-momentum $W$-algebra'') 
generated by the energy momentum tensor 
$$
T(z) = \sum_{n\in\Z} z^{-n-2} L_n\ ,
$$
whose Laurent modes satisfy the Virasoro algebra 
$$
\lb\,L_n, L_m\,\rb = (n-m)L_{n+m} + {c\over12}(n^3-n) \delta_{n+m,0} 
\eqno(2.1)$$
with {\sl central charge} $c$ taken {}from the set 
$$
c(2,q) = -\,{(3q-4)(q-3)\over q}\ ,\quad q = 2K+3\ ,\quad 
K=1,2,\ldots\ .  
\eqno(2.2)$$
The values (2.2) belong to the so-called minimal series of the Virasoro 
algebra \q{21,9}, and it is well known that for each of these central 
charges there exists a finite number of irreducible \hwr s $\h_h$ with 
highest weights $h$ in the Kac table such that those sectors together form 
a rational CFT. In our cases, the Kac table consists of the $K+1$ different 
{\sl highest weights} 
$$
h_i = -{i(q-2-i)\over 2q}\ ,\quad i=0,\ldots,K\ .
\eqno(2.3)$$
The distinguished feature of the charges and weights of the minimal models 
is the presence of two infinite chains of singular vectors in the associated
Verma modules \q{21}, and this property will be the background 
for the construction of path representations in the next subsection. 
\sn
Among the minimal models, the sub-series (2.2,3) may be characterized as 
consisting of all the theories in which the energy-momentum $W$-algebra does 
not admit an extension by additional bosonic or (para-)fermionic Virasoro 
primary fields. More precisely, they do not contain abelian sectors 
or ``simple currents'' \q{57}. 
This can be seen {}from the {\sl fusion rules} of the \ctq\  minimal models, 
which we will simply borrow {}from \q{8} and take as part of our input 
data; nevertheless, we will describe a procedure how to calculate the 
fusion rules in section 3.2.   
\sn
Let us for the moment denote the irreducible sectors  of the \ctq\ minimal 
model, $q=2K+3$, by the symbols $\phi_i$, and composition (fusion) of two 
such sectors by $\phi_i\times\phi_j$, $i,j\in I_q:=\{0, \ldots, K\}$. Then 
the representation of the observable algebra associated to 
$\phi_i\times\phi_j$ decomposes into irreducibles 
$\phi_i\times\phi_j = \sum_k N_{ij}^k\, \phi_k$ 
as 
$$
\phi_i\times\phi_j = \sum_{l=0}^{{\rm min}(i,j)} \phi_{\lbrace
|i-j|+2l\rbrace} 
\eqno(2.4)$$ 
where the abbreviation 
$$
\lbrace m \rbrace \,:=\ \cases{ &$m\quad$ if $0\leq m \leq K$\cr
&$2K+1-m\quad$ if $m>K\,$\cr}
\eqno(2.5)$$ 
was used. Note that the number of terms on the rhs of (2.4) is 
$$
m_{ij} := {\rm min}\,(i+1,j+1)\ .
\eqno(2.6)$$
The fusion rules (2.4) can be regarded as kind of ``orbifold'' of 
the multiplication table of SU(2) representations, with the sector
$\phi_1$ playing the role of the fundamental representation: 
It obeys 
$$\eqalign{
&\phi_1\times\phi_0 = \phi_1\ , \cr
&\phi_1\times\phi_i\, = \phi_{i-1}+\phi_{i+1}\quad{\rm for}\quad 0<i<K\ ,\cr
&\phi_1\times\phi_K = \phi_{K-1}+\phi_K\ .
\cr}$$
Thus, its {\sl fusion graph} -- the graph with one node $i$ per sector $i$ 
and $N_{1i}^j$ lines {}from node $i$ to node $j$ -- has a tadpole form, 
with the only loop attached to node $K$. This graph arises {}from the 
Dynkin diagram $A_{2K}$ (also familiar as the fusion graph of the fundamental 
representation sector in SU(2) WZW models) by dividing out the
$\Z_2$-symmetry.  \hfbr
\noindent The abelian ring generated by the representations $\phi_i$, 
or equivalently by the fusion matrices $N_i$ with 
$$
\bigl( N_i \bigr)_{jk} := N_{ij}^k \ ,
$$
is a {\sl polynomial fusion ring} in the sense of \q{17}: It is spanned 
by $\phi_1$ and $\check p_i(\phi_1)$, where $\check p_i(x)$ is the $i\,$th
\v Cebyshev polynomial, see also \q{40}. 
\sn 
Other distinguished sectors of the \ctq\ models are the
vacuum sector $\phi_0$ with fusion rules $\phi_0\times\phi_i=\phi_i$, 
and the sector $\phi_K$ corresponding to the representation with the 
{\sl minimal conformal dimension} of all the values listed in (2.3), 
namely $h_K = -{K(K+1)\over2(2K+3)}\,$. The fusion graph 
of $\phi_K$, denoted $\g_q$, is described by the connectivity matrix 
$$
C_q = \pmatrix{ 0&0&\cdots&0&1\cr
                0&0&\cdots&1&1\cr  
                \vdots&&\oddots&&\vdots\cr 
                0&1&\cdots&&1\cr 
                1&1&\cdots&&1\cr}
\eqno(2.7)$$
which is simply the symmetric $(K+1)\times(K+1)$ fusion matrix of $\phi_K$:
$$
\bigl( C_q \bigr)_{ij} = \bigl( N_K \bigr)_{ij} \ .
$$
The graphs $\g_q\,$, see Figure 1 for examples,  will play a surprising 
role in the following. 
\bn\bn
{\bf 2.2 Path representations and global observable algebra}
\bn
Consider some \hwr\ $\h_h$ of the Virasoro algebra at central charge $c$ with 
highest weight vector $|h\rangle$. It is well known that there exists a 
surjective homomorphism of Virasoro representations {}from the Verma module 
$\V_h$ (with the same highest weight $h$ and central charge $c$) onto $\h_h$. 
Therefore, the images of the Poincar\'e-Birkhoff-Witt (PBW) vectors (which 
are a natural basis of $\V_h$) 
$$
L_{m_1}\ldots L_{m_n} |h\rangle 
\eqno(2.8)
$$
where $m_1\leq\ldots\leq m_n<0$ and $n\geq 0$, form a {\sl spanning system}
of $\h_h$. 
If $\h_h$ is an irreducible representation occurring in the minimal models, 
it contains null-vectors and the set (2.8) is linearly 
dependent. {}From the Kac determinant \q{21}, the energy levels of the
null-states are known, and there are also recursive formulas \q{9,7} 
which in principle allow to compute them  explicitly. Following a different 
route, Feigin, Nakanishi and Ooguri succeeded in singling out an linearly 
independent subset of (2.8) for $c$ and $h$ as in (2.2,3). A closer look 
at the construction given in \q{22} shows that the main fact to be 
exploited is simply that the vacuum vector is cyclic and separating for the 
local observable algebras. This implies that the non-trivial null-vectors of 
the vacuum representation $h=0$ at $c=\ctq$ re-appear as local 
(point-like) null-fields which 
generate the ``{\sl annihilating ideal}$\,$'' \q{22} of the energy-momentum 
$W$-algebra. The annihilating ideal is represented (by zero) in all the other 
sectors as well, which produces null-vectors there.
\sn
In the vacuum Verma module $\V_0$ with central charge $c(2,q)$, $q=2K+3$, 
there is the trivial singular vector $L_{-1}|0\rangle$ and a second 
one $|s^{(q)}\rangle$ at energy $q-1$, see \q{21}, which can be 
written as 
$$
|s^{(q)}\rangle = \lim_{z\to 0} N^{(q)}(z) |0\rangle = 
N^{(q)}_{1-q} |0\rangle
$$
for a certain primary null-field $N^{(q)}(z)$ of conformal dimension $q-1$. 
Using the notations of \q{47,12}, we may give a recursive formula for the 
normal ordered products $N^{(q)}(z)$ 
$$
N^{(5)} = {\cal N}(T,T)\ ,\quad\quad N^{(q)} = {\cal N}(T,N^{(q-2)})\ ,
\eqno(2.9)$$
see also \q{22}; the normal ordering prescription denoted by ${\cal
N}(\cdot,\cdot)$ involves usual normal ordering and additional corrections 
which render the resulting field quasi-primary. For our purposes, the 
precise expression for the Laurent modes of $N^{(q)}(z)$ will be inessential, 
we only need the familiar part 
$$
N^{(q)}_n = \sum_{ {m_1,\ldots, m_{K+1} \atop m_1+\ldots+m_{K+1} = n}} \!\!
{\bf :}\,L_{m_1}\ldots L_{m_{K+1}}\,{\bf :}\  +\, \ldots 
\eqno(2.10)$$
where the dots indicate the correction terms -- which are polynomials in 
the $L_m$ of {\sl degree at most} $K$. \hfbr
\noindent {}From the vacuum Verma module, we can pass to the irreducible 
vacuum representation by quotienting out the (intersecting) sub-Verma modules 
built over the singular vectors $|s^{(q)}\rangle$ and $L_{-1}|0\rangle$. 
Translating into the space of fields, this means that we have to set equal 
zero all descendants of the null-field $N^{(q)}(z)$, i.e.\ the whole 
annihilating ideal. (The equation $L_{-1}|0\rangle=0$ translates into 
$\partial_z\one =0$ and gives no information.) This means, in particular, 
that 
$$
N_n^{(q)} |v\rangle = 0
\eqno(2.11)$$
for any mode of the null-field acting on any vector $|v\rangle$ in any of 
the representations $\h_i$ with highest weight $h_i$ (2.3). Taking 
$|v\rangle$ to be one of the highest weight vectors $|h_i\rangle$, each 
of the eqs.\ (2.11) leads to linear relations among the PBW spanning 
system (2.8). Moreover, since $N^{(q)}(z)$ is primary and generates the 
full annihilating ideal, it is also clear that all linear dependencies 
in $\h_i$ arise {}from (2.11), or {}from equations 
$\phi_m N_n^{(q)} |v\rangle=0$ with some other field
$\phi(z)$ in the energy-momentum $W$-algebra. We can,  
therefore, determine a linearly independent subset of (2.8) {}from the 
null-field modes -- provided we do the book-keeping correctly. This 
combinatorial problem can be solved by introducing a suitable 
{\sl lexicographic ordering} of the PBW vectors: We refer to \q{22} 
for the details and mention only that one of the ordering criteria is 
the length $k$ of the monomial $L_{m_1}\ldots L_{m_k}$; this should make 
it plausible that formula (2.10) for $N^{(q)}_n$ is accurate enough for 
our purposes. Taking everything together, Feigin, Nakanishi and 
Ooguri prove at the following statement:
\mn
{\bf Proposition 2.1} \q{22}\quad Let $\h_i$ be an irreducible \hwr\ of the 
\ctq\ minimal model, $q=2K+3$, with highest weight $h_i$ as in (2.3), and let 
$|h_i\rangle$ denote the  highest weight vector. Then $\h_i$ has a basis
consisting of those of the vectors 
$$ 
L_{m_1}\ldots L_{m_n}|h_i\rangle
$$ 
with $m_1\leq\ldots\leq m_n<0$ and $n\geq 0$ that satisfy the 
``{\sl difference two condition}'' 
$$
m_{l+K}-m_l \geq 2 
\eqno(2.12)$$
for $1\leq l \leq n-K$  as well as the ``{\sl initial condition}'' 
$$
\sharp \{ m_l =1\} \leq i \ .
\eqno(2.13)$$
\mn
The difference two condition (2.12) controls whether monomials in the 
Virasoro generators are ``too densely packed'': All PBW vectors that 
contain (sub-)arrays also appearing in the expansion (2.10) of null-field 
modes are removed {}from the basis. Whereas (2.12) characterizes the \ctq\ 
model as a whole, the initial condition (2.13), stating that the mode 
$L_{-1}$ appears at most $i$ times in the independent PBW monomials
belonging to $\h_i$, allows to distinguish the irreducible 
representations {}from each other. 
\mn
Let us now derive path representations {}from the FNO bases. We first 
notice that the PBW vectors (2.8), with $h$ being one of the $h_i$ in 
the list (2.3), can be uniquely expressed as {\sl sequences} of positive 
integers via the identification 
$$
L_{-M}^{\,a_M}\ldots L_{-2}^{\,a_2}L_{-1}^{\,a_1} |h_i\rangle \longmapsto 
(a_0(i); a_1, a_2, \ldots, a_M, 0, \ldots)\ ;
\eqno(2.14)$$
compared to (2.8), we have rewritten multiple equal modes as powers 
$L_{-m}^{\,a_m}$ and have also filled in the gaps (i.e.\ $a_m=0$ is 
possible). The ``dummy variable'' $a_0(i)$ is used to number the different 
sectors $h_i$. After this reformulation, the difference two condition (2.12) 
takes the simple form 
$$
a_m + a_{m+1} \leq K \quad\hbox{for all}\ m>0 \ ,
\eqno(2.15)$$
which implies that 
$$
0 \leq a_m \leq K\ ;
\eqno(2.16)$$
the initial condition (2.13) of course means that 
$$
a_1 \leq i \ .
\eqno(2.17)$$
We can get rid of this extra inequality by choosing the numbering $a_0$ 
of  sectors appropriately: Set 
$$
a_0(i) = K-i 
$$
in eq.\ (2.14), then the initial condition is nothing but the difference 
two condition (2.15) extended to $m=0$. 
\sn
We have accomplished a one-to-one map {}from the FNO basis vectors, 
Proposition 2.1, to sequences $(a_m)_{m\geq0}$ of integers 
$a_m\in\{0,\ldots,K\}$ which are subject to condition (2.16) for all 
$m\geq0$. In the next step, we encode these restrictions into the 
following  {\sl labeled graph}: It consists of $K+1$ nodes which we 
denote by $i$ running {}from 0 to $K$. The node $i$ carries the label 
$$
l(i) := K-i\ .
\eqno(2.18)$$
The graph contains (precisely) one edge {}from  node $i$ to node $j$ if 
the associated labels satisfy 
$$
l(i) + l(j) \leq K\ ,
\eqno(2.19)$$
and none otherwise. This means that there are $i+1$ unoriented 
edges emanating {}from the node $i$. As it happens, the graph 
we have described is just the {\sl fusion graph} $\g_q$ 
of the minimal dimension field $\phi_K$, with its connectivity 
matrix  given in eq.\ (2.7). 
\sn                     Now consider 
a {\sl path} on $\g_q$, i.e.\  a sequence $(i_m)_{m\geq0}$ where each 
$i_m$ is a node of $\g_q$ such that there is an edge in $\g_q$ {}from the 
node $i_m$ to the node $i_{m+1}$ for all $m\geq0$. Since the 
labeling (2.18) is unambiguous, each path of $\g_q$ is in one-to-one 
correspondence  to a sequence of labels $(l(i_m))_{m\geq0}$ which satisfy 
$0\leq l(i_m)\leq K$ and $l(i_m)+l(i_{m+1}) \leq K$ -- and these are 
precisely the requirements (2.15,16) on the sequences of $a_m$ discussed 
above. Therefore, we have established a connection {}from the FNO bases of 
the irreducible modules of the \ctq\ model to paths over the fusion 
graph $\g_q$. \hfbr
\noindent Since the monomials occurring in the PBW system (2.8) are finite, 
we also have to restrict to finite paths (to finite label sequences)  
-- or to paths  stabilizing at the node $K$, i.e.\ paths such 
that there exists an $M \gg 0$ with 
$$
i_m = K \quad\hbox{for all}\ m>M
\eqno(2.20)$$
($l(i_m)=0$ for all $m>M$). In the following we will often speak of 
{\sl finite paths} (of arbitrary length) when we mean infinite paths 
subject to the ``tail condition'' (2.20). \hfbr
\noindent The initial condition (2.17), on the other hand, really 
becomes a simple initial condition for paths: The FNO basis vectors 
of the irreducible representation $\h_j$ correspond to the (finite) 
paths starting {}from node $j$ (i.e.\ $i_0=j$). We denote the complex 
vector space generated by those paths by $\p_j$.  
\sn
Finally, our construction allows us to introduce an action of the 
energy operator $L_0$ on the path spaces $\p_i$. For each  
path $|p\rangle=(i_m)_{m\geq0}\in\p_i$, let $(l_m)_{m\geq0}$ 
be the associated sequence of labels. (Note that the associated sequences 
exist only for the paths themselves, not for arbitrary linear combinations.)
We define an ``energy operator'' $L_0^{\cal G}$ on $\p_i$ by declaring 
each path to be an eigenvector of $L_0^{\cal G}\,$: 
$$
L_0^{\cal G}\, |p\rangle = \Bigl(h_i+ \sum_{m\geq0} m\, l_m\Bigr)
\cdot|p\rangle\ ;
\eqno(2.21)$$
the constant $h_i$ is purely conventional, but with this choice 
we recover precisely the energy values in the \ctq\ representations. 
We summarize our results in  the following statement: 
\mn
{\bf Proposition 2.2}\quad Let $\g_q$ be the (labeled) fusion graph of the 
minimal dimension field $\phi_K$ in the \ctq\ model, $q=2K+3$, defined by 
the connectivity matrix (2.7) and with labels as in eq.\ (2.18). For 
$i=0,\ldots,K$, denote by $\p_i$ the vector space generated by all finite 
paths on $\g_q$ which start at node $i$. Then $\p_i$ and the irreducible 
\hwr\ $\h_i$ of the \ctq\ model are isomorphic as $\Z$-graded spaces, 
$$
\p_i \cong_{L_0} \h_i\ ,
$$
where the $\Z$-gradings are given by the usual energy operator $L_0$ on 
$\h_i$ and by the $L_0^{\cal G}$ action (2.21) on $\p_i$. 
\mn 
Let us emphasize that the path representations discussed here are 
not be confused with the path spaces underlying certain RSOS models,
see e.g.\ \q{6,50}. In particular, those paths have infinite 
tails that contribute non-trivially to the energy and even 
to the sector identification. They also do not admit the 
physical interpretation in terms of quasi-particles which 
will be discussed below.  
\mn
Up to now, we have worked with unbounded Laurent modes of point-like 
fields. If we wish, we can view the previous constructions as a 
discussion of Virasoro algebra representation theory and now 
forget all the details. Then, we 
define the path spaces $\p_i$ over the graphs $\g_q$ as the 
state spaces of the \ctq\ model. Together with the energy operator 
$L_0^{\cal G}$, which governs the ``time evolution'' of the models, 
they contain all the {\sl invariant information} 
that can be extracted {}from the FNO basis. 
\sn
In particular, we no longer need to identify ``elementary'' paths 
over $\g_q$ (i.e.\ true paths as opposed to linear combinations thereof) 
with states in the PBW system. The apparent advantage of this identification 
is that it would provide a concrete action of the Virasoro generators $L_n$ 
on the path spaces. This ``PBW type action'', however, can hardly be cast 
into a closed form and bears no relation to the path structure. In order 
to obtain useful formulas for the Virasoro action, one has to apply more 
refined methods \q{56} based on a novel interpretation of paths, which 
will be sketched in section 2.3 below. 
\sn                              {}From the point of view 
of algebraic QFT, the unbounded Virasoro generators are not of prime 
importance, anyway. Instead, one would like to construct a covariant net of
local observables $\bigl( \a(I),\ I\subset S^1 \bigr)$. 
As was mentioned in the introduction, we are not able to give a local
description of the \ctq\ models, yet, but the path picture for the 
irreducible modules at least provides a natural algebra of bounded 
operators which we can view as the {\sl global observable algebra}, 
more precisely the ``universal'' \q{24} observable algebra $\a\equiv
\a_{\rm univ}$ generated by the $\a(I)$: 
Let $\a_i = \pi_i(\a)$ denote the global (universal) observable 
algebra acting in the representation space $\h_i$ (or $\p_i$). With the 
elements of $\a_i$, we can map any state in $\h_i$ (or path in $\p_i$) 
into any other, therefore we take $\a_i$ to be the {\sl path algebra} 
associated with $\p_i\,$. 
\sn
To give a precise definition, we introduce some more notation. Let 
$\p\n_{i,j}$ be the space of paths of length $n$ over $\g_q$ which 
run {}from node $i$ to node $j$, i.e.\ 
$$
\p\n_{i,j}=\bigl\{\,|p\rangle=(i_0,\ldots,i_n)\,\big\vert\,\hbox{$i_m$\  
joined with\ $i_{m+1}$ on $\g_q$,}\ i_0=i,\ i_n=j\,
\bigr\}_{\C}\ .
\eqno(2.22)$$
The finite-dimensional algebra 
$\a\n_{i,j}$ is linearly generated by so-called {\sl strings}
$|p\rangle\langle q|$, pairs of elementary paths $|p\rangle, 
|q\rangle \in \p\n_{i,j}$  which are multiplied like matrix units,  
$$
|p\rangle\langle q|\cdot|\tilde p\rangle\langle \tilde q| = 
\delta_{q,\tilde p} \, |p\rangle\langle \tilde q|
$$
where $\delta_{q,\tilde p}$ is the Kronecker symbol. Strings act on 
paths in an analogous way, therefore 
$\a\n_{i,j}={\rm End}\,(\p\n_{i,j})$. We can endow $\p\n_{i,j}$ with a
scalar product making (elementary) paths pairwise
orthonormal; then $\a\n_{i,j}$ becomes a ${}^*$-algebra such that 
$\bigl(|p\rangle\langle q|\bigr)^*=|q\rangle\langle p|$. Denote by 
$\a_i\n$ the multi-matrix algebra $\a_i\n := \bigoplus_j \a\n_{i,j}$.
\sn
Note that this standard scalar product on the path spaces does not 
directly reproduce the Shapovalov form on the irreducible modules 
via the original identification of PBW vectors with paths. But, as 
explained, this identification involved certain choices with no 
invariant meaning. Since the (invariant) $L_0$ eigenspaces are 
finite-dimensional, it is in principle possible to define a new (pseudo) 
scalar product on the spaces of fixed energy so as to reproduce 
the Shapovalov form, but a closed formula for all energy levels is not 
known. In this context, the quasi-particle interpretation 
of paths seems to be useful, see \q{56} for further comments. 
\sn
If the graph $\g_q$ contains an edge {}from the node $j$ to 
the node $k$, we can embed the path space $\p\n_{i,j}$ into 
$\p^{(n+1)}_{i,k}$ by {\sl concatenation} of this edge to elementary paths; 
we denote the corresponding linear map by 
$$
c^j_k\,:\ \p\n_{i,j} \lra  \p^{(n+1)}_{i,k}\ ;
\eqno(2.23)$$
note that $c^j_k$ is independent of the starting node $i$. 
Concatenation induces an embedding of the simple factor $\a\n_{i,j}$ of 
$\a\n_{i}$ into $\a^{(n+1)}_{i,k}$ whenever $j$ and $k$ are joined by an 
edge of $\g_q$, i.e.\ whenever $\bigl( C_q\bigr)_{jk}=1$, see (2.7). In 
other words, $C_q$ is the {\sl embedding matrix} of the {\sl Bratteli 
diagram} associated to the injection 
$$
\Phi\n\,:\ \a\n_i \lra \a^{(n+1)}_i 
\eqno(2.24)$$
induced by (2.23). In turn, this diagram determines $\Phi\n$ 
up to inner isomorphism in $\a^{(n+1)}_i$, see e.g.\ \q{11,35}. 
\mn
{\bf Definition 2.3}\quad The global path observable algebra of the \ctq\ 
minimal model is $\a=\a_0\oplus\ldots\oplus\a_K$ where $\a_i$ is the 
$C^*$-closure of the inductive limit of the system $\bigl(\a\n_i,\Phi\n\bigr)$
with embeddings (2.24) induced by concatenation of paths. 
\mn
By construction, each generator $L_n$ of the Virasoro algebra, restricted 
to finite energy subspaces of 
the representation $\h_i$, can be expressed by elements of $\a_i\,$; 
in particular, the ``conformal Hamiltonian'' $L_0^{\cal G}$ of (2.21) 
is ``affiliated'' to \a\ in the sense that all its spectral 
projections are contained in \a. Instead of $\a_i$, we will sometimes write 
$\a_i=\pi_i(\a) \equiv {\rm pr}_i(\a)$. 
\sn
Each $\a_i$ is an {\sl AF-algebra}, i.e.\ an ``approximately
finite-dimensional'' algebra, see e.g.\ \q{11,35}. The most convenient way to 
picture AF-algebras (up to isomorphism) is by the (infinite) Bratteli diagram 
associated to the ``tower'' $\bigl(\a\n_i,\Phi\n\bigr)$. This is 
an infinite graph, subdivided into {\sl floors} which correspond to the 
finite-dimensional sub-algebras $\a\n_i$, $n=0,1,\ldots\,$. 
The $n\,$th floor consists of as many nodes as $\a\n_i$ has simple 
factors (in our case: one node on the zeroth, $K+1$ nodes on any other 
floor), and the nodes are labeled by the sizes $m$ of the factors 
$M_m(\C)$. In addition, the Bratteli diagram 
contains lines {}from each floor to the consecutive one, which fix the 
isomorphism class of the embedding $\Phi\n$ of $\a\n_i$ into $\a^{(n+1)}_i$; 
in our case, the lines are just the edges between nodes of $\g_q$, only now 
``source'' and ``range'' of an edge are viewed as belonging to different  
floors. An example is shown in Figure 2. 
\sn
Since the maps (2.24) are unital, the labeling of floors $n$ with $n\geq1$ 
follows {}from the dimensions of the factors of $\a^{(0)}_i$ and the 
structure of the (unlabeled) Bratteli diagram. The latter is the same 
for all $\a_i$, $i=0,\ldots, K$, for $q=2K+3$ fixed, except for the floors 
zero and one: It depends on the starting node $i$ which other nodes can be 
reached on the first floor; among those, however, one always finds the node
$K$, which in turn is connected to any other node of $\g_q$, thus the second 
floor already contains $K+1$ nodes. 
\sn
It is clear {}from the description that the infinite Bratteli diagram 
$\b_{q,i}$ associated to the algebra $\a_i$ simply displays all the 
infinite paths on $\g_q$ with starting node $i$ -- which is why AF-algebras 
constructed in this particular way are also called path algebras \q{49}. \hfbr
\noindent In later sections, Bratteli diagrams (finite or infinite ones) will 
provide effective tools when we are interested in certain 
statements on algebras or homomorphisms thereof only up to isomorphism. 
One such statement follows immediately {}from the observation 
that, for fixed $q$, all the $\b_{q,i}$ look alike 
except for the first few floors: This implies that the algebras $\a_i$ are
mutually {\sl stably isomorphic}, i.e.\ their infinite matrix amplifications 
$M_{\infty}(\a_i)$ are isomorphic -- see e.g.\ \q{11}.   \hfbr
\noindent Another fact which is obvious {}from the shape of the Bratteli 
diagrams is that all the $\a_i$ have trivial center. Therefore, the 
center of the global path observable algebra \a\ is $\C^{K+1}$ -- in 
accordance with the general theorems on the universal observable 
algebra $\a_{\rm univ}$ proven in \q{24}.  \
\bn\bn
{\bf 2.3 Quasi-particle interpretation of paths}
\bn
We have seen that our reformulation of the \ctq\ highest weight modules 
as path spaces gives a neat picture of these CFTs, with a 
great deal of information encoded in one labeled graph, which moreover 
is the fusion graph of a distinguished sector of the theory. While in the 
remainder of this paper, we are mainly interested in the specific 
consequence that the path representations provide us with an operator 
algebraic description of global features of the \ctq\ models, we here 
want to comment on other aspects of the path structure. We will give 
a natural interpretation of paths in terms of {\sl quasi-particles} and 
use the decomposition of the modules $\h_i$ into sub-sectors of fixed 
quasi-particle numbers to compute the {\sl characters} 
$$
\chi^K_i(q) = {\rm tr}_{{\cal H}_i}\, q^{L_0-{c\over24}} 
\eqno(2.25)$$
of the \ctq\ model with $q=2K+3$. (We hope there will be no confusion between 
the formal variable $q$ in (2.25) and the $q$ that labels the our conformal 
models.) It turns out that the quasi-particle picture of path spaces, 
details of which were first discussed in \q{56}, leads to particular 
expressions for $\chi^K_i(q)$ that reveal interesting facts about 
the physics of the \ctq\ models. 
\mn
For simplicity, let us first concentrate on the example of the $c(2,5)$ 
model, also known as the Lee-Yang edge singularity; this CFT 
describes a special critical point in the phase diagram of the Ising 
model with (complex) magnetic field \q{16}. The theory contains two 
irreducible modules with highest weights $h_0=0$ and $h_1=-{1\over5}$, 
which can be realized on the two-node tadpole graph $\g_5$, see 
Figure 1, with connectivity matrix 
$$
C_5 = \pmatrix{0&1\cr1&1\cr}\ .
$$
The label sequences $(l_m)_{m\geq0}$ associated to the paths over 
$\g_5$ are sequences of 0's and 1's satisfying 
$$
l_m = 1 \quad\Longrightarrow\quad l_{m+1}=0 \ ,
\eqno(2.26)$$
the ``tail condition'' $l_m=0$ for ``large'' $m$, and the initial 
conditions $l_0=1$ or $l_0=0$ for the vacuum or for the $h=-{1\over5}$ 
representation, resp.  
\sn
Here, we could directly interpret each sequence as a list of possible 
states of a quasi-particle, with a ``1'' in $m\,$th position indicating 
that the $m\,$th state is occupied; then (2.26) would be a generalized Pauli 
principle (``more exclusive'' than the usual one) imposed on the
quasi-particles. However, in the higher \ctq\ models, it turns out that a 
slight change of perspective is to be preferred. Namely, we identify a 
quasi-particle of the $c(2,5)$ with 
the basic length two segment $(1\,0)$ of a sequence. Then (2.26) becomes 
the ordinary Pauli principle, simply prohibiting that two quasi-particles 
``overlap''.  Except for that restriction (and the initial conditions 
on the sequences) the quasi-particles behave like 
free particles with a ``dispersion relation'' given by the energy operator 
$L_0^{\cal G}$ eq.\ (2.21). 
\sn
Let us use this to compute the energies of all $n$-quasi-particle states, 
in the $h=-{1\over5}$ representation, say. The state with the minimal 
energy corresponds to the sequence 
$$
s^{(n)}_0 = (0; 1,0,1,0,\ldots,1,0,0,\ldots)
$$
where the last ``1'' is at the $(2n-1)\,$st position; because of (2.26), the
quasi-particles cannot be packed more densely. 
$s^{(n)}_0$ has energy $E_0(n) = h_1 + 1 + 3 + 
\ldots + 2n-1 = h_1 + n^2$.    \hfbr
\noindent   The simplest excitations of this $n$-quasi-particle 
ground state are obtained by shifting the last segment ``10'' to the
right, step by step, and filling in ``0''s. The resulting states have 
energies $E_0(n)+1,\ E_0(n)+2$, etc. 
If we excite the last two quasi-particles simultaneously (by shifting 
the last segment ``1010'' to the right), we obtain states 
with energies  $E_0(n)+2,\ E_0(n)+4$, etc. The same can be done with 
$3,\ 4, \ldots, n$ quasi-particles. If we combine several of such shifts, 
we can generate all $n$-quasi-particle states {}from the ground state
$s^{(n)}_0$, and thus, the total contribution of the $n$-quasi-particle 
sector $\h_1^{(n)}$ to the character is 
$$\eqalign{
{\rm tr}_{{\cal H}_1^{(n)}}\, &q^{L_0-{c\over24}}\cr 
&\phantom{xx}= q^{h_1-{c\over24}} \cdot q^{n^2}\cdot\bigl(1+q+q^2+\ldots\bigr)
\cdot\bigl( 1+q^2+q^4+\ldots\bigr)\cdots \bigl( 1+q^n+q^{2n}+\ldots\bigr)\ ;
\cr}$$
the product occurs since the different excitations are independent  
of each other. 
\sn
The total state space $\h_1$ can be decomposed with respect to the 
quasi-particle number, 
$$
\h_1 = \bigoplus_{n\geq0} \h_1^{(n)}\ ,
$$
which immediately leads to the following {\sl sum form} for the character 
of the $h_1=-{1\over5}$ representation in the $c(2,5)$ minimal model:
$$
\chi^1_1(q) 
= q^{h_1-{c\over24}} \sum_{n\geq0} {q^{n^2}\over (q)_n} 
$$ 
where we have used the abbreviation $(q)_n := (1-q)(1-q^2)\cdots(1-q^n)$. 
For the vacuum module, corresponding to label sequences with $l_0=1$, 
the same procedure leads to the formula 
$$
\chi^1_0(q) 
= q^{h_0-{c\over24}} \sum_{n\geq0} {q^{n^2+n}\over (q)_n} \ .
$$ 
\sn
For the other \ctq\ models, one may proceed in a similar fashion. Again, the 
Hilbert spaces (or the path spaces) can be decomposed with respect to the 
quasi-particle content of the states, but now $K$ different (and independent) 
species of quasi-particles are present;  thus, 
$$
\h_i = \bigoplus_{n_1, \ldots, n_K} \h_i^{(n_1, \ldots, n_K)} 
\eqno(2.27)$$
in the \ctq\ model with $q=2K+3$. The correspondence between quasi-particles 
and basic segments of paths (i.e.\ label sequences) is as follows: Since as
in the $c(2,5)$ case, we want to interpret the difference two condition as an 
(ordinary) exclusion principle for quasi-particle, the latter must 
correspond to patterns of length two. Only now, each quasi-particle can 
occur in different ``shapes''. The ``lightest'' particle is given by the 
segment $(1\,0)$ as before, the second lightest occurs in the two forms 
$(2\,0)$ or $(1\,1)$, and so forth up to the ``heaviest'' quasi-particle 
which is the class of segments $(K\,0)$, $(K-1\,1),\ \ldots\,,\ (1\,K-1)$. 
Expressions like ``lightest'' for the moment just refer to the dispersion 
law (2.21).  Besides respecting the exclusion principle dictated by the 
difference two condition, this identification of quasi-particles with 
{\sl classes} of patterns of length two ensures that each quasi-particle 
can be excited in energy steps of one. 
\sn
It is amusing to look 
at an example (with $K\geq 2$) where a quasi-particle of the second lightest 
species upon excitation ``moves through'' one of the lightest type: 
$$
(0;2,0,1,0,\ldots)\ \hookrightarrow\ 
(0;1,1,1,0,\ldots)\ \hookrightarrow\ 
(0;1,0,2,0,\ldots)\ \hookrightarrow\ 
(0;1,0,1,1,\ldots)\ \hookrightarrow\ \ldots 
$$
Each arrow indicates that the total energy of the configuration increases 
by one unit, without changing the quasi-particle content; in the first step, 
the heavier quasi-particle changes its appearance {}from $(2\,0)$ to $(1\,1)$; 
after the second step, it has passed the lighter quasi-particle, which then 
has ``jumped'' to the left -- a phenomenon reminding of the ``time shift''
in soliton scattering. 
\sn
This sketch already suggests how to implement the Virasoro modes $L_n$ on 
the path spaces in a way which takes the path structure into account. 
In \q{56}, formulas for the su(1,1) generators $L_{\pm1}$ acting in the 
vacuum module have been constructed, following the guiding principle that 
these operators should leave the decomposition (2.27) of $\h_i$ invariant. 
(In contrast, the $L_n$ one would obtain directly {}from the PBW vectors  do 
not respect the quasi-particle numbers.) To state the precise formulas for
$L_{\pm1}$, and also to rigorously prove that the quasi-particles of 
different species can be excited independently of each other, is slightly 
technical, and we refer to \q{56} for the complete analysis. 
\sn                
Once these results are established, it is again straightforward to derive sum 
forms for the characters of the \ctq\ model, $q=2K+3$:   
$$ 
\chi^K_i(q) = q^{h_i - {c\over24}}
\sum_{n_1,\ldots,n_K\geq0} {q^{N_1^2+\cdots+N_K^2+N_{i+1}+
\cdots+N_K } \over (q)_{n_1}\cdots(q)_{n_K} }  
\eqno(2.28)$$
where $N_i:=n_i+\ldots+n_K$. The denominators show up because the different 
quasi-particles can be excited independently, the $q$-exponent in the 
numerator is the minimal energy of a configuration with $n_l$ quasi-particles
of species $l$,\ $l=1, \ldots, K$. 
\sn 
One can compare these formulas to 
the well-known Rocha-Caridi expressions for the characters of 
Virasoro minimal models, which follow directly {}from the Feigin-Fuchs 
results on the chain of singular vectors. If one applies the Jacobi 
triple product identity to the Rocha-Caridi characters, see \q{41}, one 
obtains the {\sl product form} 
\def\modbeda{{l\not\equiv0,\pm(i+1)\,({\rm mod}\,2K+3)\atop{l>0}}}
$$ 
\chi^K_i(q) =\ q^{h_i - {c\over24}}\! \prod_{\modbeda} (1-q^l)^{-1}\ . 
\eqno(2.29)$$
Equating (2.28) and (2.29) yields 
combinatorial identities known as {\sl Andrews-Gordon identities}, 
see e.g.\ \q{5}. In the special case of $K=1$, they reduce to the 
famous Rogers-Ramanujan identities. 
\mn
The surprising feature of the character sum formulas (2.28) is their 
relation to non-conformal models.  In the first place, 
expressions of the same type naturally appear in the theory of 
{\sl 1-dimensional quantum spin chains}. This was first shown in \q{39}, 
where partition sums of such chains were calculated {}from the Bethe 
Ansatz.   Kedem and McCoy also were the first to realize that the excitation 
spectrum of the chains can be interpreted in terms of quasi-particles, 
having specific dispersion relations and obeying generalized Pauli 
principles. \hfbr 
\noindent  Since 1-dimensional quantum spin chains are essentially 
equivalent to 2-dimensional statistical models on the lattice, 
the results of \q{56} on path space representations of $L_{\pm1}$ are 
also relevant as an approach towards a Virasoro action on the lattice. 
\sn
Conformal field theories arise in the continuum limit of 2-dimensional 
lattice models (or 1-dimensional quantum spin chains) at the critical 
point. Alternatively, they can appear as scaling limits of {\sl 2-dimensional 
massive QFTs} -- in particular of integrable field theories which have been 
obtained as perturbations of CFTs \q{66}. It turns out that the conformal 
characters (2.28) carry information on such perturbations as well: \hfbr
\noindent The $q$-exponent, giving the ground state energy of a 
quasi-particle configuration with prescribed 
${\tt n} := (n_1,\ldots,n_K)$, can be written as a quadratic form 
${\tt n}^{\rm t}M_q{\tt n}+ { m}_{q,i}^{\rm t}{\tt n}$ with an 
integer $K\,\times\,K$ matrix $M_q$ and integer $K$-vectors ${m}_{q,i}\,$. 
Whereas $M_q$ is ``universal'' within the \ctq\ model, ${m}_{q,i}$
also depends on the sector. \hfbr 
\noindent In our cases, $M_q$ is the inverse of the Cartan matrix of the 
$K$-node tadpole graph. The associated connectivity matrix has a 
Perron-Frobenius eigenvector $v_q$,  and the astonishing fact is that the 
ratios of the entries of $v_q$ coincide with the mass ratios of the 
particles present in the so-called $\phi_{1,3}$-perturbation of the \ctq\ 
minimal model, see \q{25}. In this sense, the expressions ``lightest'' and 
``heaviest'' quasi-particle used above receive a literal meaning. 
\sn 
Note that such ``coincidences'' occur for other sum expressions of conformal 
characters, too: By now, sum formulas for the characters of many conformal 
coset models have been found (without reference to path representations), 
see e.g.\ \q{61,38,10}, and also \q{41} for a discussion $W$-algebra 
extensions of certain minimal models.  Most remarkably, there exist two 
different sum forms for the characters of the Ising models, involving 
the inverse of the Cartan matrix of $A_1$ or of $E_8$, respectively. 
On the other hand, the two possible massive perturbations of the Ising CFT 
have either one or eight massive particles, in the latter case with 
mass ratios given by the Perron-Frobenius vector of the incidence 
matrix of the $E_8$ graph \q{66}.   
\mn
In summary, we have observed that the quasi-particle structure of the CFT 
highest weight representations and the associated sum forms of the 
conformal characters reveal certain aspects of non-conformal deformations 
(lattice models or massive QFTs) of the CFT. Although the precise 
relationship remains to be worked out, we may at least conclude 
that the path representations of the \ctq\ models are not just 
an artifact of our constructions, but do indeed have deeper 
physical meaning. 
\bn\bn
\leftline{\bf 3. The quantum symmetry algebra}
\bn 
For our construction of global amplimorphisms for the \ctq\ 
models, we will make use of an action of a quantum symmetry 
algebra (QSA) on the path representations, which 
allows to determine covariant field multiplets. 
In section 3.1, we collect some general statements on 
``quantum symmetries'' of low-dimensional QFTs. We do, however, 
in no way attempt to give a complete account of this subject, 
which has been an area of intense research during the last decade. 
We recommend e.g.\ \q{29,58} as sources where to find 
details and references to the historical development. 
Likewise, nothing new will be added to the general theory of 
quantum symmetries -- except for the remark that for so-called 
quasi-rational CFTs there exists a canonical semi-simple 
QSA, a fact that is somewhat contrary to the prevailing opinion. 
Section 3.2 is devoted to these matters, but it can be skipped if one 
is merely interested in the QSA action on the path spaces, which is set up  
in section 3.3. 
\bn\bn
{\bf 3.1 General remarks on quantum symmetries}
\bn
Local quantum field theories on a low-dimensional space-time are
interesting especially because of their superselection structure. Their 
statistics is governed by the braid group in contrast to the permutation 
group symmetry of QFTs in higher dimensions. The statistical data of a QFT 
define a representation category, which is a ``rigid braided (or symmetric) 
monoidal category with unit'', see e.g.\ \q{46,29}. A natural 
question to ask is whether this category is equivalent to the representation 
category of some group or algebra, which then could be regarded as the 
internal symmetry group (or algebra), the ``global gauge group'', of the 
model. For local QFTs in space-time dimension $\geq 3$, this problem has 
been settled by the famous duality theorem of Doplicher and Roberts \q{19} 
with the result that in this situation the statistical properties always 
``come {}from'' a compact Lie group. 
\mn
In two dimensions, and for charges localized in space-like cones in $2+1$ 
dimensions, the situation is not quite as clear yet. Research 
essentially focuses on three different variants of quantum symmetry algebras: 
\sn
One is given by {\sl quantum groups} in the sense of deformations of the
universal enveloping algebras of ordinary Lie algebras.  In certain theories, 
e.g.\ the WZW models, they arise rather naturally \q{4,1}, and they have the 
further virtue that they are closely related to lattice discretizations of 
CFTs, see e.g.\ \q{51}. It is, however, rather unlikely that quantum groups 
cover all kinds of statistics that can arise in conformal models. In 
addition, there is the unpleasant feature that  in the cases most relevant 
for rational CFTs, namely when the deformation parameter is a root of 
unity, indecomposable representations show up, the significance of 
which remains to be uncovered. 
\sn
A more speculative, and more spectacular, concept of quantum symmetry 
uses {\sl Ocneanu's string algebras} or {\sl ``paragroups''} \q{49}. They occur  
naturally in the general DHR framework as intertwiner algebras associated 
to the endomorphisms of a local QFT \q{24,28,54}. It is clear that these 
string algebras contain all information on the statistical properties of 
a theory, but a ``naive'' implementation of string algebras as ``global gauge 
symmetries'' of a QFT leads to huge total Hilbert spaces and enormous field 
algebras, as was shown in \q{53}. The ideas in \q{24,59} suggest that an 
appropriate use of string algebras as symmetry algebras requires to give up 
the clear-cut division between space-time and internal symmetries. Such a 
``mixing'' of symmetries seems to be realized in our \ctq\ models, since 
as $C^*$-algebras, the intertwiner algebras (internal symmetries) are 
simply path algebras associated to the fusion graphs of the QFT, and thus 
they are of the same kind as our path observable algebra (space-time 
symmetries). It should be interesting to pursue this relation further, but  
at present, we are interested in quantum symmetries as a mere practical 
tool and will therefore not resort to string symmetry algebras either.   
\sn
Finally, a comparatively modest approach to quantum symmetry is to rely on 
{\sl semi-simple ${}^*$-algebras} with additional structures making
them into {\sl weak quasi-triangular quasi-Hopf algebras} \q{45}. It has 
been shown in \q{58} that under certain standard assumptions such a 
semi-simple QSA always exists, in particular for rational CFTs. One 
could argue that this type of QSA does not always arise in a natural way, 
and that it may be difficult to compute the extra structure they are 
supplied with. Even more problematical, the constructions known so far 
did not single out one specific semi-simple QSA for a given model, not 
even up to isomorphism. Instead, if $i\in I$ labels the sectors of a 
rational CFT, say, with fusion rules $N_{ij}^k$, then each algebra 
$$
{\tt g} \cong \bigoplus_{i\in I} M_{n_i}(\C) 
\eqno(3.1)$$
can be viewed as a quantum symmetry of the CFT as soon as the integers $n_i$ 
satisfy the inequalities \q{58}
$$
n_i\,n_j \geq \sum_{k\in I} N_{ij}^k\,n_k\ .
\eqno(3.2)$$
We will, however, show in subsection 3.2 that for a wide class of CFTs the 
results of \q{48} yield a canonical set of such integers $n_i$. 
Since, moreover, in our special path setting the (canonical) semi-simple 
QSA leads to a very natural (and ``slim'') field algebra, and since we 
will not need the more complicated additional structures on {\tt g} for 
our purposes, it is this third concept of quantum symmetry that will be used 
in the following. 
\mn
Let us just sketch the main data of semi-simple QSAs {\tt g} and how they 
are implemented into the space of states of a CFT. Complete definitions and 
proofs can be found in \q{58}.   \hfbr
\noindent {\tt g} is a matrix algebra as in (3.1) with 
$n_0=1$ for the vacuum sector. Consequently, the 
projection onto the factor $M_{n_0}(\C)$, $\varepsilon \equiv {\rm pr}_0\,: 
{\tt g} \lra \C$, is the natural candidate for the {\sl co-unit} of {\tt g}; 
it has to be checked whether this $\varepsilon$ satisfies the correct 
relations.  \hfbr 
\noindent    Since the representation 
category of {\tt g} must ``mimic'' the braided tensor category associated to 
the CFT, the QSA {\tt g} has to be endowed with a {\sl co-product} 
$\Delta\,:\  {\tt g}\lra {\tt g}\otimes {\tt g}$ which ``reproduces'' the 
fusion rules of the CFT. The simplest way to make this requirement 
precise is to look at the  (two-floor) Bratteli diagram of $\Delta$ viewed as 
an algebra homomorphism: The diagram must contain $N_{ij}^k$ 
lines {}from the factor ${\tt g}_k \equiv e_k\cdot{\tt g}$ of ${\tt g}$ to the 
factor ${\tt g}_i\otimes {\tt g}_j$ of ${\tt g}\otimes {\tt g}$. Here, $e_i$
are the minimal central projections of {\tt g}, i.e.\  $e_i\cdot{\tt g} \cong 
M_{n_i}(\C)$. The formulation with the help of Bratteli diagrams has the 
additional advantage to make the freedom of so-called {\sl ``twists''} \q{20} 
explicit: $\Delta$ is fixed only up to conjugation with a unitary in 
${\tt g}\otimes {\tt g}$. Of course, the other data of {\tt g} have to be 
changed accordingly when $\Delta$ is twisted, since there are compatibility 
conditions; e.g.\ co-unit and co-product must obey 
$(\varepsilon \otimes\,{\rm id})\circ\Delta= {\rm id} = 
({\rm id}\,\otimes\varepsilon)\circ \Delta$. \hfbr
\noindent Since $N_{0j}^k = \delta_{kj}$ for all $k$, the co-product is always
injective, but it need {\sl not} be a {\sl unital} embedding. Obviously, 
$\Delta$ is unital iff (3.2) holds as an equality for all $i, j \in I$. \hfbr
\noindent Similarly, the co-product in general is not co-associative (it
cannot be if it is non-unital) but only  quasi-co-associative, i.e.\  
there exists a (quasi-invertible) {\sl re-associator} 
$\varphi\in {\tt g}\otimes{\tt g}\otimes{\tt g}$ intertwining   
$(\Delta\otimes\,{\rm id})\circ\Delta$ and 
$({\rm id}\,\otimes\Delta)\circ \Delta$.    \hfbr
\noindent   The braid group statistics of a low-dimensional QFT is 
reflected by the existence of an {\sl R-matrix} $R\in {\tt g}\otimes {\tt g}$ 
which is an intertwiner between $\Delta$ and $\Delta'$, the co-product with 
tensor factors interchanged. 
\sn
We have ignored the {\sl antipode} $S\,:\ {\tt g}\lra {\tt g}\,$, a 
$\C$-linear anti-automorphism of {\tt g} which translates charge 
conjugation of sectors into the QSA, and have also not discussed important 
compatibility conditions like the so-called {\sl pentagon} and 
{\sl hexagon identities}, involving $\Delta$, $\varphi$ and $R$. 
Sometimes, existence of a representation of the modular group
on {\tt g} is also assumed, see e.g.\ \q{32,62}. \hfbr 
\noindent Since we will not use those structures, we again refer to \q{58} 
for further details and merely recall that given a rational CFT with fusion 
rules $N_{ij}^k$ and a set of integers $n_i$, $i\in I$, $n_0=1$, satisfying 
(3.2), then one can solve all the constraints and obtain a weak 
quasi-triangular quasi-Hopf algebra as a QSA which reproduces  
the representation category of the CFT. 
\sn
The QSA is implemented as follows: One forms an enlarged Hilbert space 
$\h^{\rm tot}$ in which each irreducible representation space $\h_i$, 
$i\in I$, of the observable algebra occurs with multiplicity $n_i$, i.e.\ 
$$
\h^{\rm tot} = \bigoplus_{i\in I} \h_i \otimes V_i \ ,\quad\quad
{\rm dim}\,V_i = n_i\ .
\eqno(3.3)$$
Each multiplicity space $V_i$ carries an irreducible representation of 
{\tt g} which is equivalent to the defining representation $\tau_i \equiv 
{\rm pr}_i$ on $\C^{n_i}$ -- assuming that {\tt g} is directly given as 
matrix algebra. Thus, $\h^{\rm tot}$ carries a representation of {\tt g}, 
which will be denoted by $U$. 
\sn 
What we have sketched is already a special realization of the QSA within a 
QFT, of the form constructed in \q{58}. There it was shown that 
$\h^{\rm tot}$ furthermore carries a {\sl field algebra} \f\ (with a net 
structure inherited {}from \a), which can be decomposed into {\sl fields  
multiplets} transforming covariantly under the {\tt g}-action. The observable 
algebra \a\ is recovered as the fixed point algebra of \f. The Hamiltonian 
of the QFT also commutes with the {\tt g}-action. Moreover, the local 
braid relations of the field operators can be written in terms of the 
R-matrix of {\tt g}, and in this way one gains complete control of the braid 
group representations associated with the sectors of a low-dimensional QFT. 
\sn
The covariance properties of field multiplets will be an important part 
of our construction of global amplimorphisms of the observable algebra. 
They are formulated in eq.\ (4.10) below, in a slightly different form than
in \q{58}. Beyond that, we will only use information encoded in the 
co-product of the QSA, and of course the representation $U$ of {\tt g} 
on $\h^{\rm tot}$. 
\mn
First of all, we have to choose an appropriate matrix algebra 
${\tt g}_{(q)}$ as a QSA for each of our \ctq\ minimal models, i.e.\ 
we have to fix the sizes $n_i$ in eq.\ (3.1). We claim that the choice 
$$
n_i = i+1 
\eqno(3.4)$$
is one possibility; in order to show this, we prove the following: 
\sn
{\bf Lemma 3.1}\quad The integers $n_i = i+1$, $i=0,\ldots,K$ 
satisfy the inequalities 
$$
n_i\, n_j \geq \sum_k N_{ij}^k\, n_k \ ,
$$
where $N_{ij}^k$ are the fusion rules of the \ctq\ model, $q=2K+3$,  
as listed in eqs.\ (2.4,5). 
\sn
\kapp Suppose that $i\geq j$. Then eqs.\ (2.4,5) give 
$$\eqalign{
\sum_k N_{ij}^k\, n_k &= n_{i-j} + n_{\lbrace i-j+2 \rbrace} + 
\ldots + n_{\lbrace i+j\rbrace}   \cr
\noalign{\vskip -9pt}
&= i-j+1\ +\ \{i-j+2\}+1\ +\ \ldots\ +\ \{i+j\}+1   \cr
&\leq (i+1)(j+1) = n_i\,n_j \ ,
\cr}$$
where in the third line we have used the estimate $\{m\}\leq m$ 
before summing up.     \hfill\qed 
\mn
Therefore, the results collected above imply that 
$$
{\tt g}_{(q)} := \C \oplus M_2(\C)\oplus \ldots \oplus M_{K+1}(\C)
\eqno(3.5)$$
can be endowed with all the structures making it into {\sl some} 
QSA of the \ctq\ model, $q=2K+3$, and we could immediately 
proceed with implementing ${\tt g}_{(q)}$ on the total Hilbert space 
$$
\h^{\rm tot}_{(q)} := \h_0 \oplus \bigl( \h_1\otimes \C^2\bigr) \oplus 
\ldots \oplus \bigl( \h_K \otimes \C^{K+1}\bigr) \ .
\eqno(3.6)$$
But before that, we would like to take the opportunity and argue that 
the matrix algebra (3.5) can indeed be called {\sl the} 
semi-simple QSA of the \ctq\ model. 
\bn\bn
{\bf 3.2 The canonical choice of sector multiplicities}
\bn
In order to explain that the choice of dimensions $n_i =i+1$ 
in eq.\ (3.4) is even a {\sl canonical} one, we will slightly 
digress and give an account 
of some results of the work \q{48}. There, the notion of 
{\sl quasi-rational CFTs} was introduced: These are models of CFT 
such that a certain factor space of each irreducible highest weight 
module is finite-dimensional. The dimensions $n_i$ of those factor 
spaces are invariants of the theory, they satisfy  the 
sub-multiplicativity relation (3.2) and can, therefore, be used as 
dimensions of the defining representation of a semi-simple quantum 
symmetry algebra of the model. We will show that the minimal models 
of our interest are quasi-rational in Nahm's sense, and that the 
dimensions of the relevant factor spaces are given by formula (3.4). 
\sn
Up to the present, the only treatment of quasi-rationality available 
is in terms of modes of unbounded quantum fields, and using the fusion 
product as a computational tool. It would be very interesting to try 
and fit the notions developed in \q{48} into the algebraic DHR framework,  
and to find out whether there is a more conceptual interpretation of the 
dimensions $n_i$, perhaps by some operator algebraic constructions. In 
view of the inequalities (3.2), it seems likely that the $n_i$  are related 
to the Jones indices $d^2_i$ associated to the sectors, since the statistical 
dimensions $d_i$ satisfy (3.2) as equalities, see e.g.\ \q{24,43}.\hfill\break
\noindent On the other hand, it is quite clear already {}from \q{48} that  
quasi-rationality is a very useful property for practical problems: 
It allows, in particular, for an algorithmic definition of fusion in a 
large class of CFTs. Below, we will add further remarks on this aspect. 
\mn
The notion of a quasi-rational representation can be introduced for 
arbitrary bosonic $W$-algebras with a finite number of generating 
fields. A rather detailed definition of $W$-algebras can be found e.g.\ 
in \q{46,12}, but for our purposes, the following sketch is sufficient:  
\w\ contains a finite set of generating fields 
$ W^s(z) = \sum_{n\in\Z} W^s_n\, z^{-n-s} $
with conformal dimensions $s\in\Z_+$, where  $z\in\C$ is the coordinate 
of left-moving fields, and the rhs gives the expansion of  $W^s(z)$ 
in terms of Laurent modes of $W^s(z)$.  Among them, 
there is a field $W^2(z)$ of dimension two, which we identify 
with the energy momentum tensor $T(z)$; all the other 
generators $W^s(z)$ are Virasoro primary.
\w\ is then linearly generated by these fields, their derivatives with 
respect to $z$,  
and (derivatives of) normal-ordered products. Note 
that one can choose a linear basis of \w\ consisting only of primary 
generators $W^s(z)$ and quasi-primary normal ordered products. 
With respect to this 
basis, \w\ is an infinite-dimensional Lie algebra, and for most of what 
follows, we may simply regard a $W$-algebra as the  
universal enveloping algebra generated by the Laurent modes of the 
fields $W^s(z)$ -- in this way avoiding a general 
discussion of normal ordered products. 
\mn 
We denote by $\w_{--}$ the linear span of modes  
$$\eqalign{
\w_{--} &:= 
\bigl\{\,\phi_n \,|\, \phi(z)\in\w,\ \phi\neq\one,\  
n\leq - {\rm dim}\,\phi\,\bigr\}
\cr
&\,= \bigl\{\, {\textstyle \oint\omega(z)\phi(z)}\,|\  \phi(z)\in\w,\ 
\phi\neq\one,\ \omega\ \hbox{a 1-form vanishing at}\ \infty\,\bigr\}\ ,
\cr}\eqno(3.7)$$
where in the second description we have used  contour integration 
around zero to project onto Laurent modes. 
\sn
{\bf Definition 3.2}\quad Let $V$ be an irreducible highest weight 
representation of a finitely generated bosonic $W$-algebra \w, 
and put $V\ua := \w_{--} V$. The representation $V$ is called 
{\sl quasi-rational} if the quotient space $V/V\ua$ is finite-dimensional. 
A CFT with finitely generated bosonic $W$-algebra 
is called {\sl quasi-rational} if all the 
irreducible \hwr s making up the chiral space of states are quasi-rational. 
\mn
Note that in the last part of the definition we do {\sl not} require that the 
CFT involves only a finite number of irreducible representations, i.e.\ 
that it is rational. On the one hand, there is the plausible conjecture 
that any rational CFT with appropriate $W$-algebra is quasi-rational.
On the other hand, there are definitely examples of non-rational 
quasi-rational theories -- which is one of the advantages of working with 
Definition 3.2. In particular, many of the $N=2$ superconformal QFT
associated to Calabi-Yau manifolds probably are non-rational, but since 
they are relevant for string theory, it is important to have tools 
which work in a larger class of CFT than just the rational ones. 
The restriction to bosonic $W$-algebras in Definition 3.2 
is not essential, see \q{34} for an extension to the fermionic case. 
\mn
Given a representation $V_i$ of a quasi-rational CFT, we use the notation 
$$
n_i = {\rm dim}\,\bigl( V_i/V_i\ua \bigr) 
\eqno(3.8)
$$
for the dimension of the factor space. Note that $n_i\geq1$ for any 
representation. To a subspace $V_i\us \subset V_i$ 
such that dim$\,V_i\us = n_i$ and $V_i\us + V_i\ua$ is dense in $V_i$, we 
will refer to as a {\sl small space} of the CFT. While the integers $n_i$ 
are invariantly defined, the small spaces are not. Often, however, there 
are natural choices in explicit computations. 
\mn
The most interesting property of the dimensions $n_i$ was also proven in
\q{48}: 
\sn
{\bf Proposition 3.3}\ \q{48}\quad Consider a CFT with $W$-algebra \w\ and 
let $V_m$, $m\in{I}$, be the collection of irreducible \w-representations
occurring in the CFT. Assume that among these there are two  quasi-rational 
\hwr s $V_i$ and $V_j$ with small space dimensions $n_i$ and $n_j$. 
Denoting the fusion rules of the theory by $N_{ij}^k$ for 
$i,j,k \in {\cal I}$, we have the following inequality: 
$$
n_i\, n_j \geq \sum_k N_{ij}^k\, n_k 
$$
\sn
{\bf Corollary 3.4}\quad The set of quasi-rational \hwr s of a $W$-algebra 
forms a sub-category (of the category of all \hwr s) which is 
closed under fusion. \hfbr 
\noindent Quasi-rational representations are semi-rational, i.e.\ the 
fusion decomposition of two quasi-rational representations contains only 
a finite number of irreducibles. \hfbr
\noindent The $n_i$ can be used as dimensions of the defining representation 
of a semi-simple quantum symmetry algebra of a quasi-rational CFT. 
\mn
To sketch the proof of Proposition 3.3, we have to use the {\sl fusion 
product} of two $W$-algebra representations. For $\alpha=1,2$, choose two 
distinct ``punctures'' $z_{\alpha}\in\C$ and ``insert'' a representation 
$V_{i_{\alpha}}$ at $z_{\alpha}$. The action of a field $\phi(z)\in\w$ on 
a vector $v_{\alpha}\in V_{i_{\alpha}}(z_{\alpha})$ can be written as a 
contour integral 
$$
\oint_{C_{\alpha}}\! \omega(z)\phi(z)\, v_{\alpha}
$$
where $\omega(z)$ is a meromorphic 1-form and $C_{\alpha}$ is a ``small'' 
contour encircling $z_{\alpha}$. The well-known fusion product 
representation of \w\ on the tensor product $V_{i_1}\otimes V_{i_2}$ is 
defined via Cauchy's formula 
$$
\oint_{C_{12}}\! \omega(z)\phi(z)\,( v_1 \otimes v_2 ) :=
\bigl(\oint_{C_1}\! \omega(z)\phi(z)\,v_1\bigr)
\otimes v_2
+ v_1 \otimes \oint_{C_2}\! \omega(z)\phi(z)\,v_2\ ,
\eqno(3.9)$$
where $C_{12}$ encircles both $z_1$ and $z_2$. Introducing a mode expansion 
on the lhs -- here, a choice is involved -- leads to a 
$z_{\alpha}$-dependent, co-product-like expression of the fusion product 
action of $\phi_n$ on $V_{i_1}\otimes V_{i_2}$ in terms of modes $\phi_m$ 
that act in each $V_{i_{\alpha}}$ separately, see 
e.g.\ \q{23,46,48}. If the action (3.9) can be diagonalized we obtain 
a decomposition into irreducibles -- hence the name ``fusion product''. 
\mn
Contour integrals may be  used to prove Proposition 3.3 in the following way. 
Consider any  vector  $v_1\otimes v_2 \in V_{i_1}\otimes V_{i_2}$ such that 
$v_{\alpha}\in V_{i_{\alpha}}$ are $L_0$-eigenvectors. Having chosen small 
spaces $V_{i_{\alpha}}\us$, we can decompose $v_2$ into a sum 
$v_2=v_2\us+v_2\ua$ with $v_2\us\in V_{i_2}\us$, $v_2\ua\in V_{i_2}\ua$. By 
definition, $v_2\ua$ is of the form $v_2\ua=\phi_{-h-n}v_2'$
for some $v_2'\in V_{i_2}$ and  some field mode $\phi_{-h-n} \in\w_{--}$, 
where $h$ is the dimension of $\phi$ and $n\geq0$.   
Using eq.\ (3.9), this can be rewritten as 
$$
v_1\otimes v_2 = \phi_{-h-n}(v_1\otimes v_2')  -  
\Bigl(\,\sum_{l\geq0} f_l(z_1-z_2) \phi_{l-h+1}\,v_1\,\Bigr)
\otimes v_2'
\eqno(3.10)$$
where $f_l(z_1-z_2)$ are certain meromorphic functions arising {}from 
the mode expansion. The first term on the rhs of (3.10) is in 
$\bigl(V_{i_1}\otimes V_{i_2}\bigr){}\ua$, and the remaining
terms have a lower $L_0$-degree than $v_1\otimes v_2$. Proceeding
inductively, and treating $v_1$ in the same way, one finally  
reaches vectors in the tensor product of the small spaces, which
proves that $V_{i_1}\otimes V_{i_2}$ is generated by vectors in 
$\bigl(V_{i_1}\otimes V_{i_2}\bigr){}\ua$ and $V_{i_1}\us \otimes V_{i_2}\us$. 
In other words, the space $V_{i_1}\us \otimes V_{i_2}\us$ contains 
a small space $\bigl(V_{i_1}\otimes V_{i_2}\bigr){}\us$ of $V_{i_1}\otimes
V_{i_2}$ endowed with the \w-action {}from the fusion product. 
By definition, this representation contains all other irreducible \hwr s 
$V_k$ with multiplicity $N_{i_1i_2}^k$ which implies the desired 
inequality.  For more details of  the proof, we refer to \q{48}. We  
would, however, like to emphasize that the statement in Proposition 3.3 is 
independent on any choice to be made when giving explicit formulas 
for the fusion product action of field modes.    
\mn
The practical use of the concept of quasi-rationality combined 
with the fusion product now becomes apparent: It allows to determine 
the fusion rules  of quasi-rational representations by a finite algorithm. 
One simply has to diagonalize the zero-mode sub-algebra $\w_0$ of \w\ on the 
finite-dimensional space $V_{i_1}\us \otimes V_{i_2}\us$ in order to obtain 
a decomposition into irreducibles.  In this procedure, however, two subtleties
are hidden. First of all, diagonalizability may fail to hold in general, 
although the known examples for such a situation are typically plagued with 
certain pathological features. The other problem involves the so-called 
{\sl spurious states}, i.e.\  vectors in the space 
$V_{i_1i_2}^{\sigma} := \bigl(V_{i_1}\us \otimes V_{i_2}\us\bigr) \cap 
\bigl(V_{i_1}\otimes V_{i_2}\bigr){}\ua$. This intersection is non-zero if the 
inequality in Proposition 3.3 becomes strict, e.g.\ in most minimal models 
of the Virasoro algebra. Fusion rules are then obtained {}from diagonalizing 
the zero mode action in the space 
$\bigl(V_{i_1}\us \otimes V_2\us\bigr)/ V_{i_1i_2}^{\sigma}$. The construction
of spurious states is not yet well understood at an abstract level, although 
in concrete examples it is usually possible to determine 
$V_{i_1i_2}^{\sigma}$  {}from the null-fields of the theory. 
\sn
As was shown in \q{48}, the fusion rules can be calculated -- up to the 
subtleties mentioned above -- on even smaller spaces: Denote by $V_i\uh$ the 
highest weight subspace (wrt $L_0$) of the module $V_i$, then one can show 
in a similar fashion as before  that both $V_{i_1}\uh\otimes V_{i_2}\us$ and  
$V_{i_1}\us\otimes V_{i_2}\uh$ contain the space 
$\bigl(V_{i_1}\otimes V_{i_2}\bigr){}\uh$. This further reduction is very 
useful in concrete applications, and often allows to avoid cumbersome 
calculations with spurious states. 
\mn
We will now show that the minimal models with central charge $c(2,q)$ 
and highest weights $h_i$ as in eqs.\ (2.2,3) are quasi-rational CFTs
and that the dimensions of their small spaces are given by the simple
formula (3.4). \hfbr
\noindent Consider a \hwr\  with highest weight 
$$
h_{r,s} \equiv h_{r,s}(p,q) = {(pr-qs)^2 - (p-q)^2 \over 4pq} 
\eqno(3.11)$$
of the Virasoro algebra at central charge $c(p,q)$, where $r,s\geq 1$ are 
integers, but we admit arbitrary $p,q\in\R_{>0}\,$. Feigin and Fuchs have 
shown that the Verma module $\V_h$ over $|h_{r,s}\rangle$ contains a 
singular vector $|v\rangle$ at level $r\cdot s$ (and maybe others, all 
at higher energy), which is of the form 
$$
|v\rangle = L_{-1}^{rs}|h_{r,s} \rangle + \alpha |v'\rangle 
\eqno(3.12)$$
where the complex number $\alpha$ depends on $r, s$ and $p/q$,  
and $|v'\rangle \in \V_h$ involves  the modes $L_{-2},\, L_{-3}, 
\ldots$ -- see 
\q{21} for more explicit statements. After passing to the irreducible 
module $\h_{r,s}$, eq.\ (3.12) implies that 
$$
L_{-1}^{rs}|h_{r,s} \rangle \in \bigl(\h_{r,s}\bigr)\ua
$$
since $L_{-2},\, L_{-3}, \ldots \in {\rm Vir}_{--}\,$, 
cf. eq.\ (3.7). On the other hand, 
one can easily convince oneself -- using, e.g., the explicit formula for 
normal ordered products in \q{47,12} -- that none of the vectors 
$|h_{r,s}\rangle,\ L_{-1}|h_{r,s}\rangle, \ldots, 
L_{-1}^{rs-1}|h_{r,s}\rangle$ can be written as 
$\phi_m|v'' \rangle$ for some $|v''\rangle\in \h_{r,s}$, some 
normal ordered product $\phi(z)$ of the energy momentum tensor, 
and $m\leq -{\rm dim}\,\phi$. Thus, for the degenerate models 
of the Virasoro algebra with highest weights $h_{r,s}(p,q)$, the 
vectors listed above can be taken as a basis of the small spaces; 
the equation 
$$
n_{r,s} \equiv {\rm dim}\, \bigl(\h_{r,s}\bigr)\us = rs 
$$
for arbitrary degenerate representations and in particular formula (3.4)
for our models follow. As a further consequence, we indeed obtain 
the sub-multiplicativity relation in Lemma 3.1 as a special case of 
Proposition 3.3, without resort to any combinatorics. 
\sn
The models with $h_{r,s}(p,q)$ as above and $r, s\in\Z_+\,$, 
$p/q \notin \Q$, are simple 
examples of non-rational quasi-rational CFTs, and of theories where the  
dimensions $n_{r,s}$ satisfy the relation of Proposition 3.3 as an equality. 
In contrast, for the minimal models $c(2,q)$, $q\geq5$ an odd integer, 
we have $n_i n_j > \sum_k N_{ij}^k n_k$ in general, which also means 
that there exist spurious states.  
\sn 
As mentioned above, this fact complicates the computation of fusion 
rules by diagonalizing the $L_0$-action on $\h_i\us\otimes \h_j\us$ 
or $\h_i\us \otimes \h_j\uh$. Nevertheless, the special cases we 
are interested in are simple enough to let us circumvent these problems, 
e.g.\ in the following three ways:  \hfbr
\noindent  First of all, we may ``perturb'' the models slightly by 
moving the central charge away {}from the minimal values $q\in\Z$
while keeping the relations (3.11) for the highest weights. Then, the 
model becomes merely degenerate, the spurious states disappear, and we 
may read off the $N_{ij}^k$ {}from the $L_0$-action in the tensor product 
of small spaces. Afterwards, we can move $c$ back to the minimal value, 
take into account the conformal grid symmetry $h_{r,s} = h_{q-r, p-s}$ 
for minimal models if necessary, and we will recover the fusion rules 
of the minimal model. In the limit $c \rightarrow c_{\rm min.mod.}$ 
some higher level vectors in the irreducible components 
of the fusion product move towards singular vectors of the minimal 
representations, thus producing the spurious states in a controlled way.\hfbr
\noindent The second possibility is to calculate the fusion rules directly 
within the minimal model $c(2,q)$, but using diagonalization of $L_0$ in 
the smaller spaces $\h_i\us \otimes \h_j\uh$. All the highest weight spaces 
are, of course, one-dimensional in pure Virasoro models. For each $q$, the 
small space $\h_1\us$ of the \hwr\ with $h_1=-{q-3\over2q}$ is 
two-dimensional, so we can obtain the fusion rules of $\h_1$ with 
$\h_k,\ k\neq 0$, {}from diagonalizing $2\times2$ matrices. But in each \ctq\ 
model, the conformal family corresponding to $\h_1$ generates the whole 
fusion ring.        \hfbr 
\noindent The most direct method, however, is to calculate the spurious 
states explicitly. In the $c(2,q)$ models, this can be done since the 
null-fields are known, see eq.\ (2.9). Their Laurent modes applied to  
highest weight states give null-vectors
which can be used to compute the spurious states via contour integration
and Cauchy's theorem. We refrain {}from giving further details here and 
rather refer to \q{48,3}. 
\mn 
In summary, we have seen that Nahm's results \q{48} indeed provide a 
{\sl canonical} choice for the isomorphism type of a {\sl semi-simple 
quantum symmetry algebra} for a quasi-rational CFT, namely 
$$
{\tt g} = \bigoplus_{i\in{\cal I}} M_{n_i} (\C) 
$$
where $n_i = {\rm dim}\,\h_i\us$ are the small space dimensions associated 
to the irreducible representations. The $n_i$ are {\sl invariants} 
of the quasi-rational CFT.  \hfbr
\noindent    One can show that for quasi-rational models  
even the other basic data of {\tt g}, namely the co-product, the R-matrix
and the re-associator, can be reconstructed explicitly: The small spaces can 
be used to define a finite-dimensional vector bundle equipped with a 
flat connection, which leads to a generalization of the Knizhnik-Zamolodchikov
equation {}from WZW models to arbitrary quasi-rational CFTs. Then, Drinfeld's
construction \q{20} can be applied to recover all data of a  weak 
quasi-triangular quasi Hopf algebra {}from the  generalized 
Knizhnik-Zamolodchikov equation \q{3}.  
\sn
The fusion rules are part of this structure, to be determined
by the diagonalization procedure sketched above. Therefore, we can in
principle delete them {}from the input data listed in section 2.1 and instead 
derive them {}from information referring only to individual \hwr s. 
\bn\bn
{\bf 3.3 Action of the quantum symmetry algebra on the path spaces}
\bn
Irrespective of whether the sector multiplicities $n_i = i+1$ chosen in 
eq.\ (3.4) are canonical or not, we will see that they lead to a particularly
natural action of the associated quantum symmetry algebra ${\tt g}_{(q)}$ 
on the ``amplified'' Hilbert space $\h^{\rm tot}_{(q)}$, see eqs.\ (3.5,6), 
of the minimal model with central charge \ctq, $q=2K+3$. The point   
is that for this special choice of $n_i$, the space $\h^{\rm tot}_{(q)}$ can 
again be represented as a path space over a Bratteli diagram $\widehat{\b}_q$ 
that has essentially the same form as the $\b_{q,i}$ underlying the 
individual spaces $\h_i$.  
\sn 
The extended Bratteli diagram  $\widehat{\b}_q$ looks as follows: Floors
are numbered $-2,\ -1,\ 0,\ 1,\ \ldots\,$; the $-2\,$nd floor consists of one
node, labeled $*$, all other floors of $K+1$ nodes labeled $0, \ldots, K$ 
as usual; the embedding matrices between floors $l$ and $l+1$ for $l\geq -1$ 
are the $(K+1)\times (K+1)$ connectivity matrices $C_q$ of the \ctq-graph 
$\g_q$, see eq.\ (2.7), whereas the matrix describing the embedding of 
floor $-2$ into floor $-1$ is simply $C_{\rm in}=(1,1,\ldots,1)^{\rm t}\in 
M_{(K+1)\times 1}(\Z)$ -- i.e.\ the node $*$ is joined to every node on 
the $-1\,$st floor by a single edge. \hfbr
\noindent Compared to the Bratteli diagrams $\b_{q,i}$ of section 2.2, 
the only new building block of $\widehat{\b}_q$ is an extremely simple,
``canonical'' one. In Figure 2, an example is shown. 
\mn
Let $\widehat{\p}{}^{(n)}_{*,i}$ denote the space of all paths on 
 $\widehat{\b}_q$ {}from the 
node $*$ on the $-2\,$nd floor to the node $i$ on the $n\,$th floor, 
for $n\geq -1$ and $i= 0, \ldots, K$. 
\sn
{\bf Lemma 3.5}\quad 
${\  {\rm dim}\,\widehat{\p}{}^{(0)}_{*,i} = n_i}$ 
\sn
\kapp We compute the dimension of this path space by applying the first two 
embedding matrices to the dimension 1 of the space of length zero paths: 
$$  
{\rm dim}\,\widehat{\p}{}^{(0)}_{*,i}=\epsilon_i^{\rm t}\,C_q\,C_{\rm in}\,1 
=  \epsilon_i^{\rm t}\, (1,2, \ldots, K+1)^{\rm t} = i+1 = n_i\ ,
$$
where $\epsilon_i$ is the $i\,$th standard unit vector, with rows 
labeled {}from 0 to $K$. \hfill\qed 
\mn
As a consequence, we obtain an ($L_0$-graded) isomorphism between the 
space $\widehat{\p}_*$ of all (finite) paths over the extended Bratteli 
diagram $\widehat{\b}_q$ and the total Hilbert space $\h^{\rm tot}_{(q)}$, 
$$
\widehat{\p}_{*}   \cong \h^{\rm tot}_{(q)} \ : 
\eqno(3.13)$$
In order to show this, we first label the paths 
in  $\widehat{\p}{}^{(0)}_{*,i}$ by $\nu$
running {}from 1 to $n_i$; then we identify a state in the $\nu\,$th copy of
$\h_i$ with a path on $\widehat{\b}_q$ which reaches node $i$ on the $0\,$th
floor along the path $|\nu\rangle$, and continues towards infinity according 
to the path space representations of each $\h_i$ introduced in section 2.2. 
\sn
With a different choice of multiplicities $n_i$, it 
would still be possible to find an extended Bratteli diagram whose associated
path space is isomorphic to $\h^{\rm tot}$, but it could be very 
different in shape {}from the one determined by the graph $\g_q$. In other 
words, the path field algebra associated to the extended Bratteli diagram, 
see below, would no longer be of the same stable isomorphism type as 
the $\a_i$.  
\mn
Let us, for one more time, refer to the ideas of section 3.2 where the 
multiplicities $n_i$ were interpreted as the dimensions of special subspaces
$\h_i\us\subset\h_i$. For degenerate Virasoro models, we could simply choose 
$$
\h_i\us = \bigl\{\, L_{-1}^m |h_i\rangle\,\big\vert\, m\geq0,\  L_{-1}^m 
|h_i\rangle\ \hbox{linearly independent of}\ L_n\h_i\ {\rm for\ all}\  
n\leq -2\,\bigr\}_{\C}\ .
$$
With this information, we may derive that $\widehat{\p}_{*}\cong
\h^{\rm tot}_{(q)}$ even without knowledge of the values of $n_i$: Recall 
that by Proposition 2.2 and with the notations of (2.22) we have  
$$
\h_i\us \ \cong \ \bigoplus_{l=0}^{K} P^{(1)}_{i,l}\ \cong\  
\bigoplus_{l=0}^{K} P^{(1)}_{l,i}\ ;
$$
the second relation holds since the graphs $\g_q$ are unoriented. 
But the last decomposition just describes the space of paths on the  
two extra floors of $\widehat{\b}_q$, ending at node $i$ on floor 0.  
Since {}from floor zero on, the extended diagram is as the diagrams $\b_{q,i}$, 
its associated path space $\widehat{\p}_{*}$ is simply $\bigoplus_i
\h_i\otimes\h_i\us$, no matter what the dimensions of the small 
spaces are. By construction, this space is our total space of states.  
\mn
Given the path representation of the total Hilbert space, the action of 
the quantum symmetry algebra is implemented in a straightforward way. For 
each fixed $q$, we denote by $\f{}^{(n)}_{*,i}$ the string algebra
over $\widehat{\b}_q$ generated by pairs of paths joining node $*$ on floor 
$-2$ to node $i$ on floor $n$. We set $\f{}^{(n)}_{*} := 
\bigoplus_i \f{}^{(n)}_{*,i}$, and we enumerate the paths 
$|\nu\rangle \in \widehat{\p}^{(0)}_{*,i}$ {}from $\nu=0,\ldots,n_i$ as before. 
\sn
{\bf Definition 3.6}\quad The quantum symmetry 
algebra ${\tt g} = \bigoplus_i M_{n_i}(\C)$ acts as an associative algebra 
on the total Hilbert space $\h^{\rm tot}$ by the representation $U$ which 
is defined on matrix units $E^i_{\mu\nu} \in M_{n_i}(\C)$ as 
$$
U\bigl(E^i_{\mu\nu}\bigr)=|\mu\rangle\langle\nu|\in\f^{(0)}_{*,i}\ ;
$$
in other words, $U({\tt g})= \f^{(0)}_{*}$. \
\mn
This natural action of {\tt g} on $\h^{\rm tot}$ is unital and faithful. The 
operators $U(a)$, $a\in{\tt g}$, flip only the first two edges of a path in 
$\widehat{\p}_*$, leaving already the node on the $0\,$th floor fixed. 
On the highest weight state $|h_i^{\mu}\rangle$ of the $\mu\,$th copy of 
$\h_i$, $\mu=1,\ldots,n_i$, they act as 
$$ U(a)  |h_i^{\mu}\rangle = \sum_{\nu=1}^{n_i} (a_i)_{\mu\nu}\,|h_i^{\nu}
\rangle\ ; 
$$
the complex numbers $(a_i)_{\mu\nu}$ are the matrix entries of the $i\,$th 
factor of $a\in{\tt g}$. As a consequence, the vacuum is ``invariant'' under 
{\tt g} in the sense that it transforms in the trivial representation given 
by the co-unit. (This fact reminds us that {\tt g} generalizes the group 
algebra rather than the Lie algebra of a global gauge group.) In particular, 
we conclude that the  {\tt g}-action has no ``observable effect'':
\mn
{\bf Proposition 3.7}\quad Let \f\ denote the path algebra generated by 
all finite strings over $\widehat{\b}_q$. The commutant of $U({\tt g})$ in 
\f\ is canonically isomorphic to the observable algebra, 
$$
U({\tt g})' \cap \f \cong \a\ .
$$
\sn
\kapp The claim is more or less obvious, and the proof will be given 
mainly in order to set up a matrix notation for the elements of \f\ 
which will be useful in later computations. The identification of elementary
strings with matrix units has already been used for finite floor 
sub-algebras; now, it will be applied to make the form of field algebra 
elements explicit up to a certain floor: For fixed $q= 2K+3$, let 
$N= n_0 + n_1+ \ldots + n_K = {(K+1)(K+2)\over2}$ be the total 
number of paths on $\widehat{\b}_q$ {}from $*$ to all the nodes on the 
zeroth floor. Then we can write any element $F\in\f$ as matrix 
$$
F = \bigl( F_{rs} \bigr)_{r,s=1}^{N}  
\eqno(3.14)$$
where $F_{rs}$ is a linear combination of strings starting at $*$, 
taking the routes $|r\rangle\langle s|$ down to the $0\,$th floor, 
and then continuing arbitrarily (such that the ends eventually meet). 
In this notation, the operators in $U({\tt g})$ have the block-diagonal   
form
$$
U(a) \in {\rm bl\,diag}\,\bigl( \C\cdot \one_{{\cal A}_0}, 
M_2(\C\cdot \one_{{\cal A}_1}), \ldots, 
M_{K+1}(\C\cdot \one_{{\cal A}_K})\,\bigr) 
$$
for all $a\in{\tt g}$, where $\one_{{\cal A}_i}$ is the unit of the 
path algebra $\a_i$. 
The commutant of $U({\tt g})$ in \f\ is then simply given by 
$$
U({\tt g})' =  {\rm bl\,diag}\,\bigl( \a_0, D_2(\a_1), \ldots, 
D_{K+1}(\a_k)\,\bigr)
\eqno(3.15)$$
where $D_i\,:\ \a \lra \a^{\oplus i},\ a \longmapsto a\oplus\cdots
\oplus a$ is the diagonal embedding; obviously, $U({\tt g})'$ is 
canonically isomorphic to \a.         \hfill\qed
\mn
In view of this proposition, we will call \f\ the (path representation of 
the) {\sl field algebra} of the \ctq-model. Of course, like our path 
observable algebra, \f\ is  a global object. As an AF-algebra, \f\ once    
more belongs to the same stable isomorphism class as all the path algebras 
associated to the path representations $\h_i$, which is clear {}from the 
form of the Bratteli diagram $\widehat{\b}_q$. Unlike the global observable 
algebra \a, however, \f\ is a simple algebra. Note also that $\f \neq 
U({\tt g}) \lor  U({\tt g})'$, but the centers of the QSA and the
observable algebra \a\ coincide on the total Hilbert space, 
$$
Z(\a) = Z(U({\tt g}))\ ,
$$
in agreement with the general theory. Compared to the huge field algebras
constructed elsewhere, our path field algebra \f\ envelops  
the path observable algebra \a\ rather tightly.  
\bn\bn
\leftline{\bf 4. Covariant field multiplets}
\bn
This section contains the main step towards global amplimorphisms of the 
\ctq\ minimal models, namely the construction of covariant field multiplets 
inside the field algebra \f. Once these are known, the amplimorphisms follow 
immediately. In order to arrive at the multiplets, the QSA is first endowed
with a co-product $\Delta$ reproducing the fusion 
rules of the \ctq\ model -- or rather with an equivalent collection of
amplimorphisms $\nu_i$, $i=0, \ldots, K$, of {\tt g}. With the help of these 
amplimorphisms, one can formulate equations on elements of \f\ which 
are to form covariant multiplets. In our cases, the special properties of the 
underlying path spaces allow for a natural solution of the
covariance conditions. 
\bn\bn
{\bf 4.1 Amplimorphisms of the QSA} 
\bn
For the moment, we treat {\tt g} as an abstract matrix algebra 
${\tt g} = \bigoplus_{i=0}^K M_{n_i}(\C)$. The very first condition {\tt g} 
has to meet in order to become the QSA of a CFT is existence of a co-product 
$\Delta\,:\ {\tt g}\lra {\tt g}\otimes {\tt g}$ which reproduces the fusion 
rules of the CFT, see section 3.1. This means  that 
the (two-floor) Bratteli diagram of the algebra homomorphism 
$\Delta\,:\ {\tt g}\lra {\tt g}\otimes {\tt g}$ is fixed, but there is 
the freedom of ``twisting'' the co-product by inner automorphisms of 
${\tt g}\otimes{\tt g}$. 
\sn
Given $\Delta\,:\ {\tt g}\lra {\tt g}\otimes {\tt g}$, we can use the 
minimal central projections $e_i$, $i=0, \ldots, K$, to introduce 
{\sl amplimorphisms} \q{62} 
$$
\nu_i\,:\ \cases{ &${\tt g} \lra M_{n_i}({\tt g}) $ \cr
\noalign{\vskip 5pt}
&$a \longmapsto \nu_i(a) := (\one \otimes e_i) \Delta(a)$ \cr}
$$
of {\tt g}; here, the isomorphism $M_{n_i}({\tt g}) \cong 
{\tt g} \otimes M_{n_i}(\C)$ has tacitly been applied. 
\sn
Vice versa, a collection of amplimorphisms 
$\nu_i\,:\ {\tt g} \lra M_{n_i}({\tt g})$ for $i=0, \ldots, K$, of the 
above type defines a co-product
$$
\Delta_{\{\nu\}}\,:\ \cases{ &${\tt g} \lra  {\tt g}\otimes {\tt g}$\cr
\nvs
&${  a \longmapsto \Delta(a) := \sum_{i=0}^K  \nu_i(a)}\  $ \cr} 
$$
-- up to inner automorphism, since again an explicit isomorphism {}from 
$M_{n_i}({\tt g})$ to ${\tt g} \otimes {\tt g}_i$, $g_i = e_i\cdot {\tt g}$, 
has to be chosen. The invariant information contained in $\Delta$ and 
in the collection of {\tt g}-amplimorphisms is, however, the same, and 
we will see  that the objects we aim at, namely the 
amplimorphisms of the observable algebra, are independent of twists of
$\Delta$. 
\sn
To work with amplimorphisms of the semi-simple QSA instead of the co-product 
was proposed by Szlachanyi and Vecsernyes, and the idea has been applied to 
$G$-spin chains and to the Ising model \q{60,62}. Amplimorphisms seem to 
be better adapted to the DHR framework, and, for some purposes, are easier 
to handle in practice. 
\mn
Our task is, given the \ctq-fusion rules $N_{ij}^k$, find amplimorphisms
$\nu_i$ of ${\tt g}$ whose Bratteli diagrams contain $N_{ij}^k$ lines {}from 
${\tt g}_k$ to $M_{n_i}({\tt g}_j)$. This is fairly easy to do,   
in fact, the main difficulties in writing down the $\nu_i$ are 
of notational type. Fix a superselection sector $i\in I_q$, and choose, 
for each $j\in I_q$, an enumeration 
$$
(ij|1), (ij|2), \ldots, (ij|m_{ij})
\eqno(4.1)$$
of the fusion results $|i-j|, \{|i-j|+2\}, \ldots, i+j$ in the decomposition
of $\phi_i\times\phi_j$,  
see eqs.\ (2.4-6); recall that $m_{ij} = {\rm min}(i,j) +1$. Using this 
notation, and the decomposition of {\tt g} and its elements with the help 
of the minimal central projections, $a_i := a\cdot e_i$, 
we define the auxiliary map 
$$
\widetilde{\nu_i}\,:\ \cases{&${\tt g} \lra M_{n_i}({\tt g}_0) \oplus 
M_{n_i}({\tt g}_1) \oplus \cdots \oplus M_{n_i}({\tt g}_K)$ \cr\nvs
&${ a \longmapsto\widetilde{\nu_i}(a)}$\cr}
\eqno(4.2)$$
by 
$$
\widetilde{\nu_i}(a) := a_i \oplus {\rm bl\,diag}\bigl( a_{(i1|1)},\ldots, 
a_{(i1|m_{i1})}, 0_{d_{i1}} \bigr) \oplus \cdots \oplus
{\rm bl\,diag}\bigl( a_{(iK|1)},\ldots, a_{(iK|m_{iK})}, 0_{d_{iK}} \bigr)\ ; 
$$
$0_{d_{ij}}$ denotes a square matrix of size 
$$
d_{ij} := n_i n_j - \sum_{l=1}^{m_{ij}} n_{(ij|l)}
\eqno(4.3)$$
with all entries equal to zero. The $d_{ij}$ are just the defects in the 
inequalities (3.2). In the terminology of section 3.2,
$d_{ij}$ is the dimensions of the ``spurious space'' in the fusion of the
representations $i$ and $j$. 
\sn
The target space of $\widetilde{\nu_i}$ is isomorphic 
to $M_{n_i}({\tt g})$, and we arrive at a true amplimorphism of the 
QSA if we choose a specific isomorphism. Since both $\bigoplus_l 
M_{n_i}({\tt g}_l)$ and $M_{n_i}({\tt g})$ embed canonically into 
$M_{n_iN}({\C})$, $N=\sum_l n_l$, we can simply use a permutation matrix
$P_{\pi} \in M_{n_iN}({\C})$ in order to ``rearrange'' the matrix 
$\widetilde{\nu_i}(a) \in  M_{n_iN}({\C})$. The 
permutation $\pi\in S_{n_iN}$ is obtained {}from 
comparing the following two labelings of the standard basis $\epsilon_s$, 
$s=1, \ldots, n_iN$ of $\C^{n_iN}$: 
\mn
{\sl Labeling I} is in terms of triples $(n_j, \sigma; \alpha)$ with 
$j = 0, \ldots, K$, $\sigma = 1, \ldots, n_j$, $\alpha = 1, \ldots, n_i$. 
Here, $\epsilon_{(n_j, \sigma; \alpha)} = \epsilon_s$ with 
$s= n_i (n_0 + \ldots + n_{j-1}) + (\alpha-1)n_j + \sigma\,$.  
\sn
{\sl Labeling II} is in terms of triples $(\alpha; n_j, \sigma)$ 
with the same range of indices. Here, 
$\epsilon_{(\alpha; n_j, \sigma)} = \epsilon_s$ with 
$s= (\alpha-1)N + (n_0 + \ldots + n_{j-1}) + \sigma\,$. 
\mn
The triples $(n_j, \sigma; \alpha)$ of labeling I provide a natural 
row-column ordering for the sub-algebra $\bigoplus_l M_{n_i}({\tt g}_l)$ 
of $M_{n_iN}({\C})$, whereas the triples $(\alpha; n_j, \sigma)$ of 
labeling II are appropriate when we consider the sub-algebra 
$M_{n_i}({\tt g})$. Accordingly, let $\pi\in S_{n_iN}$ be the 
permutation $s \longmapsto \pi(s)$ with 
$$
\pi(s) = (\alpha; n_j, \sigma) \quad{\rm for}\quad s = (n_j, \sigma; \alpha)
\eqno(4.4)$$ 
in an obvious notation, and let $P_{\pi}$ be the $n_iN \times n_iN$ 
matrix whose $s\,$th column is the standard unit vector
$\epsilon_{\pi(s)}$. 
\mn
{\bf Proposition 4.1}\quad For all $i\in I_q$, the map 
$$
\nu_i\,:\ \cases{&${\tt g} \lra M_{n_i}({\tt g})$ \cr\nvs
&${  a \longmapsto  P_{\pi}^{\rm t}\, \widetilde{\nu_i}(a)\, 
P_{\pi}}\ , $\cr}
$$
with $P_{\pi}$ determined by (4.4) as above,  is an amplimorphism of the 
QSA {\tt g} such that any associated co-product $\Delta_{\{\nu\}}$ 
reproduces the \ctq-fusion rules. 
\sn
\kapp  First, we have to show that $P_{\pi}^{\rm t}\,\widetilde{\nu_i}(a)\,
P_{\pi} \in M_{n_iN}(\C)$ is contained in the sub-algebra $M_{n_i}({\tt g})$
for all $a\in {\tt g}$. This is true since $\widetilde{\nu_i}(a)$ is 
an element of the block-diagonal sub-algebra $M_{n_i}({\tt g}_0) \oplus 
\cdots \oplus M_{n_i}({\tt g}_K)$ of $M_{n_iN}(\C)$, which means, in 
terms of labeling I, that  
$\bigl(\widetilde{\nu_i}(a) \bigr)_{(n_j,\sigma;\alpha)
(n_k,\tau; \beta)} \neq 0$ only if $j=k$;   
the same applies for $\bigl(\nu_i(a) \bigr)_{(\alpha; n_j,\sigma)
(\beta; n_k,\tau)}$, only now the counting is wrt labeling II, and 
$\nu_i(a) \in M_{n_i}({\tt g})$ follows. An explicit formula 
for the matrix elements of the amplimorphism $\nu_i$ is  
$$
\bigl(\nu_i(a) \bigr)_{(\alpha; n_j,\sigma)(\beta; n_k,\tau)} 
= \delta_{jk} \sum_{l=1}^{m_{ij}} \bigl( a_{(ij|l)} \bigr)_{f_l(
\alpha,\sigma),f_l(\beta,\tau)} 
\eqno(4.5{\rm a})$$
with 
$$
f_l(\alpha,\sigma) = \sigma + (\alpha-1)n_j - 
(n_{(ij|1)} + \ldots + n_{(ij|l-1)})\ ,
\eqno(4.5{\rm b})$$
and we use the convention that the matrix element $(a_l)_{\rho\sigma}$ 
vanishes unless $1\leq\rho,\sigma\leq n_l$. Thus, at most one of the terms 
in the sum of eq.\ (4.5a) gives a non-zero contribution. \hfbr
\noindent The assertion on the Bratteli diagram associated to the 
amplimorphism $\nu_i$ (and to $\Delta_{\{\nu\}}\,$) is enforced by the very 
construction of the
auxiliary maps $\widetilde{\nu}_i$, which have been chosen as the simplest 
possible realization of the Bratteli diagram dictated by the fusion rules. 
Both in the definition of $\widetilde{\nu}_i$ and of  $\nu_i$  
one could introduce additional twists by unitaries. \hfill\qed
\mn
Note that the amplimorphisms $\nu_i$ are ${}^*$-homomorphisms
with respect to the canonical ${}^*$-operation on complex matrix 
algebras and, more importantly, that they are all non-unital -- 
except for $\nu_0 = {\rm id}$. This non-unitality may be 
traced back to the non-zero defects $d_{ij}$ of eq.\ (4.3), but 
the rank of $\nu_i(1)$ does not depend on the specifically simple choice of 
amplimorphisms $\nu_i$. Likewise, the following important statement is one 
on the inner isomorphism class of the {\tt g}-amplimorphism:
\mn
{\bf Proposition 4.2}\quad Upon composition, the amplimorphisms 
$\nu_i$ of {\tt g} 
realize the fusion rules, i.e.\  for each pair $i,j\in I_q\,$, there 
exists a unitary $U_{ij}\in M_{n_in_j}({\tt g})$ such that 
$$
( \nu_i \circ \nu_j ) (a) = U_{ij}^*\, \bigl( \bigoplus_{k\in I_q} 
\nu_k (a)^{\oplus N_{ij}^k} \bigr)\, U_{ij} 
$$
for all $a\in{\tt g}$ and the fusion rules given in section 2.1. In   
particular, $\nu_i\circ\nu_j$ and $\nu_j\circ\nu_i$ are unitarily equivalent. 
\sn
\kapp Recall that an algebra homomorphism $\psi\,:\ A \lra B$ is 
extended to an algebra homomorphism  
$\psi\,:\ M_n(A) \lra M_n(B)$ of the amplifications 
by setting $\psi\bigl((a_{ij})_{i,j=1}^n\bigr)
= \bigl( \psi(a_{ij})\bigr)_{i,j=1}^n$. Thus, the lhs is a map {}from 
{\tt g} to $M_{n_i}(M_{n_j}({\tt g}))$, the rhs can be regarded
as an element of $M_{n_in_j}({\tt g})$ because of the basic inequality
$n_in_j \geq \sum_k N_{ij}^k n_k$. The existence of the unitaries 
$U_{ij}$ is clear since in the Bratteli diagrams associated to $\nu_i\circ
\nu_j$ and $\bigoplus_k \nu_k^{\oplus N_{ij}^k}$  
there are $\sum_k N_{im}^k N_{jk}^l$ resp.\ $\sum_k N_{ij}^k N_{km}^l$
lines {}from the factor ${\tt g}_l$ to the factor $M_{n_in_j}({\tt g}_m)\,$:
The diagrams coincide.  Note that with our simple realization of the 
amplimorphisms, the unitaries $U_{ij}$ are in fact permutation 
matrices. \hfill\qed
\bn
Let us illustrate the notions of this section in the simplest model 
of our series, namely the CFT describing the Lee-Yang edge singularity 
of the Ising model. To the $c(2,5)$ minimal model, we associate the 
QSA 
$$
{\tt g}_{(5)} = \C \oplus M_2(\C)\ . 
$$
(In the language of small spaces {}from subsection 3.2, dim$\,\h_0\us=1$ 
for the vacuum representation is a general fact, and dim$\,\h_1\us=2$
follows {}from the presence of the null-vector $(L_{-1}^2 - {2\over5}
L_{-2}) |h_1\rangle =0$ in the irreducible module $\h_1$ of the $c(2,5)$ 
theory.)  
\sn 
The total Hilbert space $\h_{(5)}^{\rm tot}=\h_0\oplus(\h_1\otimes\C^2)$ 
is identified with the path space over the 
Bratteli diagram $\widehat{\b}_{5}$, see Figure 2,  
and the global (path) field algebra is the string algebra over 
$\widehat{\b}_{5}$. The representation of ${\tt g}_{(5)}$ on 
$\h_{(5)}^{\rm tot}$ is implemented in a straightforward way following 
Dfinition 3.6. 
\mn
The amplimorphisms of ${\tt g}_{(5)}$ are also obtained easily: 
$\nu_0 = {\rm id}$ is trivial, and the fusion rule $\phi_1 \times
\phi_1 = \phi_0 + \phi_1$ with this enumeration of fusion results, 
i.e.\ $(11|1)=0,\ (11,2)=1$, yields 
$$
\widetilde{\nu_1}(a) = 
\pmatrix{ a_1^{11} & a_1^{12}\cr a_1^{21} & a_1^{22}\cr} \oplus
\pmatrix{ a_0 & 0 & 0 & 0 \cr 0& a_1^{11} & a_1^{12} & 0 \cr
          0&  a_1^{21} & a_1^{22} & 0 \cr 0 & 0 & 0 & 0 \cr} 
$$
for the auxiliary morphism $\widetilde{\nu_1}\,:\ {\tt g} \lra
M_2({\tt g}_0) \oplus M_2({\tt g}_1)$, where 
$a = a_0 \oplus \pmatrix{a_1^{11}& a_1^{12}\cr a_1^{21}&a_1^{22}\cr} \in
{\tt g}$. In the $c(2,5)$-model, the defect is $d_{11}=1$, explaining the 
zero in the lower right corner of the matrix $\widetilde{\nu_1}(a)$. \hfbr
\noindent The permutation $P_{\pi}$ which has to be applied in 
passing to an amplimorphism $\nu_1\,:\ {\tt g} \lra M_2({\tt g})$ 
amounts to slicing the matrices in $\widetilde{\nu_1}(a)$ into 
quarters and collecting them together as 
$$
\nu_1(a) = \pmatrix{\  a_1^{11} \oplus \pmatrix{a_0&0\cr 0&a_1^{11}\cr} 
 & a_1^{12} \oplus \pmatrix{0&0\cr a_1^{12}&0\cr}\  \cr  
 \phantom{\sum} &\phantom{\sum} \cr
\  a_1^{21} \oplus \pmatrix{0&a_1^{21}\cr 0&0\cr} 
 & a_1^{22} \oplus \pmatrix{a_1^{22}&0\cr 0&0\cr}\  \cr}\ .
\eqno(4.6)$$
It is straightforward to verify that $(\nu_1 \circ \nu_1)(a)$ 
and $\nu_0(a) \oplus \nu_1(a)$ are equal up to simultaneous permutation 
of rows and columns. 
\bn\bn
\leftline{\bf 4.2 Construction of covariant field multiplets}
\bn
In this subsection, we will construct multiplets of ``charged fields''
which transform covariantly under the QSA action, and which can be used to 
define (global) amplimorphisms of the path observable algebra \a. 
\sn
The charged fields associated with a sector $i\in I_q$ are operators 
$F_i = \bigl(( F_i)_{\alpha\beta} \bigr)_{\alpha,\beta=1}^{n_i} \in 
M_{n_i}(\f)$ in the $n_i$'th amplification of the path field algebra \f, 
and they are subject to the conditions 
$$\eqalignno{
a\cdot\bigl(F_i\bigr)_{\alpha\beta}  &= \bigl( F_i \cdot \nu_i(a)
\bigr)_{\alpha\beta} \quad\quad \hbox{for all}\ a\in{\tt g}\ , 
&(4.7)\cr
F_i^* \vphantom{\sum^k}F_i &= \nu_i(1) \ .
&(4.8)\cr}$$ 
In both equations, we have identified $a\in{\tt g}$ with its image $U(a)$ on 
$\h^{\rm tot}$. Eq.\ (4.7) simply expresses the {\tt g}-covariance of the 
charged multiplets, written in terms of {\tt g}-amplimorphisms, cf.\ \q{62}, 
rather than the co-product as in \q{58}.  
The second relation, where ``1'' of course is the unit 
of the QSA {\tt g}, can be viewed as a completeness condition; however, 
since the {\tt g}-amplimorphisms $\nu_i$ are i.g.\ non-unital, the field 
multiplets $F_i$ are merely partial isometries in $M_{n_i}(\f)$. 
\mn
Finding explicit solutions of (4.7) and (4.8) is the only slightly 
technical part of the constructions presented in this paper. The matrix 
notation of eq.\ (3.14) for the field operators $(F_i)_{\alpha\beta}$ will 
prove useful in this process, only now we will group the indices according 
to ``labeling II'' of the last subsection. Thus, we write 
$$
\bigl(F_i\bigr)_{\alpha\beta} = \bigl( (F_i)_{(\alpha; n_k,\sigma)
(\beta; n_j,\rho)} \bigr)_{k,\sigma; j,\rho} 
\eqno(4.9)$$
with $j,k \in I_q$, $\rho = 1,\ldots, n_j$, $\sigma=1,\ldots,n_k$. Again, 
this notation is to make the first two floors of a string in \f\ ``visible'': 
The indices $(n_j,\rho)(n_k,\sigma)$ of the $(F_i)_{\alpha\beta}\,$-entry 
indicate that it is a (linear combination of) string(s) of the form 
$| p^k_{\infty}\circ\sigma_k^*\rangle\langle p^j_{\infty}\circ \rho_j^*|$ 
where $|\rho_j^*\rangle$ is the $\rho$'th path {}from $*$ to the node $j$ on 
the 0'th floor, and $|p_{\infty}^j\rangle\in\p_j$ runs {}from $j$ on to 
infinity (with the usual tail condition); the symbol $\circ$ denotes 
concatenation of the two pieces. We will call such an element of \f\ with
indices $(n_k,\sigma)(n_j,\rho)$ a {\sl $k$-$j$-string} for short (though 
it is actually only a ``half-open'' string). \hfbr
\noindent Since we can identify states in the irreducible representations 
$\p_j$ with {\sl $j$-paths} on the extended Bratteli diagram (in obvious 
terminology), we find that the field operators carry one representation 
into another. More precisely, the matrix element (4.9) maps
the ($\rho$'th copy of the) space $\h_j$ into the ($\sigma$'th copy of the)
space $\h_k$. 
\mn
In the following, we will first deal with the covariance condition (4.7), 
which turns out to determine only the ``coarse structure'' of the operators 
$\bigl(F_i\bigr)_{\alpha\beta}$ -- i.e.\ to determine that some of the 
matrix elements (4.9) have to vanish, and some have to coincide. 
Afterwards, the completeness relation (4.8) will place non-trivial 
constraints on the non-vanishing entries in $\bigl(F_i\bigr)_{\alpha
\beta}\,$, and it is this step where the combinatorial structure of the 
path spaces becomes essential in solving the constraints. 
\mn
{\bf Proposition 4.3}\quad The field multiplet
$\bigl(F_i\bigr)_{\alpha\beta}\,$, $\alpha, \beta = 1, \ldots, n_i$,
transforms covariantly under the {\tt g}-action if and only if the matrix
elements (4.9) have the following form:
$$
(F_i)_{(\alpha; n_k,\sigma)(\beta; n_j,\rho)} = C^k_{ij;\,\alpha}\ 
\sum_{l=1}^{m_{ij}}  \delta_{(ij|l),k} \,\,
\delta_{\sigma+n_{(ij|1)}+\ldots+n_{(ij|l-1)}\,,\,\rho+(\beta-1)n_j }
$$
Here, $C^k_{ij;\,\alpha}$ denotes some $k$-$j$-string on the Bratteli diagram 
$\widehat{{\cal B}}_q\,$. In particular, the matrix elements vanish unless 
$\phi_k$ occurs in the fusion decomposition of $\phi_i\times\phi_j\,$.
\sn
\kapp One advantage of the explicit matrix notation for field algebra 
elements is that the QSA-operators $a\in{\tt g}$ commute with the entries
$C^k_{ij;\,\alpha}$ of $\bigl(F_i\bigr)_{\alpha\beta}\,$, since $a\equiv
U(a)$ acts non-trivially only between $*$ and the 0'th floor of
$\widehat{{\cal B}}_q\,$, whereas the $C^k_{ij;\,\alpha}$ are made up {}from 
half-open string starting on the 0'th floor. Therefore, we can simply treat 
$C^k_{ij;\,\alpha}$ as numerical coefficients, for the time being. \hfbr
\noindent With $F_i$ as above, let us calculate the lhs of eq.\ (4.7): 
$$\eqalign{
\bigl( a\cdot (F_i)_{\alpha\beta}\bigr)&_{(n_m,\tau)(n_j,\rho)} 
= \sum_{(n_k,\sigma)} a_{(n_m,\tau)(n_k,\sigma)}\,\,(F_i)_{(\alpha;
n_k,\sigma)(\beta;n_j,\rho)} 
\cr
&=\sum_{(n_k,\sigma)} \delta_{m,k}\,\,(a_m)_{\tau,\sigma}\,\,
C^k_{ij;\,\alpha}\  \sum_{l=1}^{m_{ij}} \delta_{(ij|l),k} \,\,
\delta_{\sigma+n_{(ij|1)}+\cdots+n_{(ij|l-1)}\,,\,\rho+(\beta-1)n_j }
\cr
&=  C^m_{ij;\,\alpha}\   \sum_{l=1}^{m_{ij}} \delta_{(ij|l),m} \,\,
(a_m)_{\tau, f_l(\beta,\rho)}
\cr}$$
with $f_l(\beta,\rho)=\rho+(\beta-1)n_j-n_{(ij|1)}-\ldots-n_{(ij|l-1)}$ {}from 
(4.5b). For the rhs of (4.7), we obtain 
$$\eqalign{
\bigl( F_i \nu_i(a) \bigr)&_{(\alpha; n_m,\tau)(\beta; n_j,\rho)} 
= \sum_{(\gamma; n_k,\sigma)} C^m_{ik;\,\alpha}\ 
 \sum_{l=1}^{m_{ij}} \delta_{(ij|l),k} \,\, 
\delta_{\tau+n_{(ik|1)}+\ldots+n_{(ik|l-1)}\,,\,\sigma+(\gamma-1)n_k } \cr
&\phantom{xxxxxxxxxxxxxxxxxxxxxx}\times\, \delta_{k,j} \sum_{l'=1}^{m_{ik}}
\bigl(a_{(ik|l')}\bigr)_{f_{l'}(\gamma,\sigma), f_{l'}(\beta,\rho)}     \cr
&= C^m_{ij;\,\alpha}\   \sum_{l,l'=1}^{m_{ij}}  \delta_{(ij|l),m}\,\,
\bigl(a_{(ik|l')}\bigr)_{g_{l,l'}(\tau),f_{l'}(\beta,\rho)}\ ; 
\cr}$$
here, we have introduced the shorthand 
$g_{l,l'}(\tau) = \tau+ n_{(ik|1)}+\ldots+n_{(ik|l-1)}-
n_{(ik|1)}-\ldots-n_{(ik|l'-1)}$ for the first index.  
Now recall our convention that
$(a_m)_{\rho,\sigma}=0$ unless $1\leq\rho,\sigma\leq n_m\,$, and also that we 
have once and for all fixed an enumeration of the fusion results $(ij|1),
\ldots, (ij|m_{ij})$ in eq.\ (4.1). Put together, this means that the 
last expression contains an implicit Kronecker symbol setting $l=l'$ and, 
therefore, it agrees with the lhs of (4.7) calculated before.   \hfbr
Given the formula (4.5) for the {\tt g}-amplimorphisms, one can also show 
that the multiplets $F_i$ must have the form given in the proposition in 
order to solve (4.7). One may e.g.\  insert matrix units for $a\in{\tt g}$
to achieve complete decoupling of all equations, and the only difficulty 
is to keep track of the indices. Since this is slightly tedious, and 
since later we will not aim at that the most general solution for the 
coefficients $C_{ij;\,\alpha}^k\,$, we omit details of the ``only if''  
part of the proof.        \hfill\qed
\mn
Let us now turn to the completeness relation. We first prove the following 
intermediate result, which does not yet involve the special structure of 
our path spaces. 
\mn
{\bf Proposition 4.4}\quad  The field multiplets $F_i$ solve the 
completeness relation (4.8) if and only if the coefficients
$C_{ij;\,\alpha}^k$ of Proposition 4.3 satisfy 
$$
\sum_{\alpha=1}^{n_i}  \bigl(C_{im;\,\alpha}^k\bigr)^*  C_{ij;\,\alpha}^k
= \delta_{N_{ij}^k,1}\,\delta_{m,j}\,\,\one_{{\cal P}_j}\ ,
\eqno(4.10)$$ 
where $\one_{{\cal P}_j}$ is the identity operator on the path space $\p_j$ 
of section 2, viewed as subspace of $\widehat{\p}_{(q)}\,$. 
\sn
\kapp Assume that (4.10) holds, and insert the formula for $F_i$ {}from 
Proposition 4.3 into the lhs of (4.8); this  yields 
$$\eqalign{
\bigl( F_i^* F_i\bigr)&_{(\alpha; n_m,\tau)(\beta; n_j,\rho)} 
= \sum_{(\gamma; n_k,\sigma)} \bigl((F_i)_{(\gamma; n_k,\sigma)
(\alpha;n_m,\tau)}\bigr)^*
(F_i)_{(\gamma; n_k,\sigma)(\beta; n_j,\rho)} 
\cr
&= \sum_{(\gamma;n_k,\sigma)}\bigl(C_{im;\,\gamma}^k\bigr)^*C_{ij;\,\gamma}^k
\sum_{l'=1}^{m_{im}} \sum_{l=1}^{m_{ij}}  
\delta_{(im|l'),k}\, \delta_{(ij|l),k}\,\cr
&\phantom{xxxxxx}\times
\delta_{\sigma+n_{(im|1)}+\ldots+n_{(im|l'-1)}\,,\,\tau+(\alpha-1)n_m }\,
\delta_{\sigma+n_{(ij|1)}+\ldots+n_{(ij|l-1)}\,,\,\rho+(\beta-1)n_j }
\cr
&= \one_{{\cal P}_j} \delta_{m,j} \sum_{(n_k,\sigma)}\sum_{l=1}^{m_{ij}} 
\delta_{(ij|l),k}\,\,\delta_{\tau+(\alpha-1)n_j\,,\,\rho+(\beta-1)n_j}\,\,
\delta_{\sigma+n_{(ij|1)}+\ldots+n_{(ij|l-1)}\,,\,\tau+(\alpha-1)n_j } \ .
\cr}$$
We have again used the fact that the enumeration of the fusion results was 
fixed, which enforces $l=l'$ above. Since the range of both $\tau$ and 
$\rho$ is $1, \ldots, n_j$, the last but one Kronecker symbol implies 
$\tau=\rho$ {\sl and} $\alpha=\beta$. Thus, $F_i^* F_i$ is diagonal, 
$$
\bigl( F_i^* F_i\bigr)_{(\alpha; n_m,\tau)(\beta; n_j,\rho)} 
= \delta_{\alpha,\beta}\,\delta_{m,j}\,\delta_{\tau,\rho} \,
\Theta_{ij}(\alpha,\tau)\,\,\one_{{\cal P}_j} 
$$
with a ``cutoff factor''
$$
\Theta_{ij}(\alpha,\tau)  := \cases{ 1 &if $\sum_{k} N^k_{ij}n_{k} 
\geq \tau+(\alpha-1)n_j\,$, \cr
0 &otherwise.\cr}
$$
The rhs of (4.8) is only a special case of (4.5), 
$$
\bigl(\nu_i(1)\bigr)_{(\alpha; n_m,\tau)(\beta; n_j,\rho)} 
= \one_{{\cal P}_j} \,\delta_{m,j} \sum_{l=1}^{m_{ij}} 
\bigl(1_{(ij|l)}\bigr)_{f_l(\alpha,\tau), f_l(\beta,\sigma)}\ ,
$$
which we have multiplied by the unit operator on $\p_j$ as we are actually 
working with
$U(\nu_i(1))$ acting on $\widehat{\p}_{(q)}\,$. Since $1\in{\tt g}$ is a 
diagonal matrix, the elements above vanish unless $f_l(\alpha,\tau)= 
f_l(\beta,\sigma)$, i.e.\ unless $\tau=\rho$ {\sl and} $\alpha=\beta$, 
and a closer look at the ``defect'' of the non-unital amplimorphism $\nu_i$ 
shows that the cutoff factor $\Theta_{ij}(\alpha,\tau)$ appears as
well.   \hfbr
\noindent  Proving the reverse direction is easier after having a 
closer look at the structure of the matrix elements of $F_i$ in 
Proposition 4.3: Note that the Kronecker symbols are independent of 
$\alpha$ and that, for fixed $\alpha$, there is at most one non-zero 
entry in each column $(\beta; n_j,\rho)$. Thus, the entries in 
$F^*_i F_i$ are precisely of the form
$\sum_{\alpha}\bigl(C_{im;\,\alpha}^k\bigr)^* C_{ij;\,\alpha}^k$, and 
condition (4.10) follows {}from the matrix structure of $\nu_i(1)$. \hfill\qed
\mn
The actual task is to construct (half-open) strings $C_{ij;\,\gamma}^k$ which 
satisfy the relations (4.10). This will be done with the help of embeddings 
of path spaces, mapping elementary paths to elementary paths. First, we need 
a combinatorial lemma comparing the sizes of certain path spaces. 
\mn
{\bf Lemma 4.5}\quad As in eq.\ (2.22), let $\p^{(2)}_{k,m}$ be the space of 
paths of length 2 on $\g_q$ which start {}from node $k$ on the $0\,$th 
floor and end at node $m$ on the $2\,$nd floor, $k,m\in I_q$. With the 
\ctq\ fusion rules $N_{ij}^k$ and the sector multiplicities $n_i=i+1$, 
the following estimate holds for all $i,k,m\in I_q\,$: 
$$
n_i\,{\rm dim}\,\p^{(2)}_{k,m} \geq \sum_{j\in I_q} N_{ij}^k 
\,{\rm dim}\,\p^{(2)}_{j,m}
$$
\sn
\kapp As in the proof of Lemma 3.5, we express ${\rm dim}\,\p^{(2)}_{j,m}$ 
through the embedding matrix $C_q$ and standard unit vectors; thus, the lhs is
$$
n_i\,{\rm dim}\,\p^{(2)}_{k,m} = n_i\, \epsilon_m^{\rm t} C_q^2 \epsilon_k = 
n_i\, (N_K^2)_{mk} = n_i\,( N_0+N_1+\ldots+N_K)_{mk}\ .
$$
Here, the fact that $\g_q$ is just the fusion graph of the minimal dimension 
field $\phi_K$ of the \ctq-model is very convenient: The last equality is 
the fusion rule $\phi_K\times\phi_K$. For the rhs, we compute
in the same fashion that 
$$\eqalign{
 \sum_{k} N_{ij}^k \,{\rm dim}\,\p^{(2)}_{j,m} &= \sum_k N_{ij}^k(N_K^2)_{mj}
= (N_K^2 N_i)_{mk} \cr
&=\bigl( (i+1)(N_K + \ldots + N_i) + i N_{i-1}+\ldots+2 N_1+N_0\bigr)_{mk}\ .
\cr}
$$
The last step follows {}from applying the fusion rule 
$\phi_K\times\phi_i = \phi_K+\phi_{K-1} +\ldots+ \phi_{K-i}$ twice, see
section 2.1.  Subtracting the rhs {}from the lhs, we obtain a matrix whose 
elements are all non-negative if and only if $n_i\geq i+1$.    \hfill\qed
\mn         {}From the proof 
of this lemma, we learn as a by-product that our choice of sector 
multiplicities is indeed the {\sl minimal} one such that the dimension 
estimate holds true -- and, as a consequence, such that the 
construction of the $C_{ij:\,\alpha}^k$ to be given below is possible. 
This seems remarkable since up to now the special values $n_i=i+1$ were 
distinguished only on the general grounds of section 3.2, whereas the 
basic inequalities $n_i n_j \geq \sum_k N_{ij}^k n_k$ can in general 
be fulfilled with some of the multiplicities taken smaller than $i+1$. 
\mn
Below, the following generalizations of Lemma 4.5 will be useful:
\sn
{\bf Corollary 4.6}\quad For all $n\geq2$, and for all $i,k,m\in I_q$,
we have 
$$
n_i\,{\rm dim}\,\p^{(n)}_{k,m} \geq \sum_{j\in I_q} N_{ij}^k 
\,{\rm dim}\,\p^{(n)}_{j,m}\ .
$$
Furthermore, if the sector indices are such that $k+m\geq K$, the dimension
inequality also holds for $n=1$. 
\sn
\kapp The first claim is true because all path spaces are based on the same 
fusion graph $\g_q$, so for $n\geq2$ we obtain the dimensions of the path 
spaces $\p^{(n)}$ when applying $C^{n-2}_q$ to those of $\p^{(2)}$; this  
does not spoil the estimate in the previous lemma.   The case $n=1$ is only 
special as far as some of the spaces $\p^{(1)}_{k,m}$ are empty: These are 
precisely those with $k+m<K$, as follows {}from the form of $\g_q$. \hfill\qed
\mn
After these preparations, we are ready to construct strings
$C_{ij;\,\alpha}^k$ with the property (4.10). Lemma 4.5 guarantees that 
for all $i,j,k,m\in I_q$ there exist injective homomorphisms 
$$
\iota_{i,\,m}^{k\ (2)}\,:\ \bigoplus_j \bigl( \p^{(2)}_{j,m} \bigr)^{\oplus
N_{ij}^k}   \lra  \bigl(\p^{(2)}_{k,m} \bigr)^{\oplus n_i} 
\eqno(4.11)$$
of path spaces of length 2, which leave the endpoints (here the node $m$)
fixed. We arrange the injections in such a way that elementary paths are 
mapped to elementary paths. This allows us to extend $\iota^{(2)}$ to longer
paths simply by requiring compatibility with concatenations $c^l_m$, see 
eq.\ (2.23): Given a collection of maps $\iota_{i,\,l}^{k\ (n)}$ 
for some $n\geq2$, we define $\iota_{i,\,m}^{k\ (n+1)}$ by
$$
\iota_{i,\,m}^{k\ (n+1)}\bigl( c^l_m(p) \bigr) =
c^l_m\bigl(\iota_{i,\,l}^{k\ (n)}(p) \bigr)
\eqno(4.12)$$
for all elementary paths $|p\rangle\in\bigoplus_k\bigl(\p^{(n)}_{j,l} 
\bigr)^{\oplus N_{ij}^k}$.  Moreover, the second part of Corollary 4.6 
states that injections $\iota^{(1)}$ can already be defined for paths of 
length 1 at least in some cases. Among those are the first edge of the 
highest weight path (wrt the $L^{\cal G}_0$ action (2.21) in each $\p_i$, 
i.e.\ the edge $(i\rightarrow K)$; we choose $\iota^{k\,(1)}_i$ such as 
to map this edge to $(k\rightarrow K)$ -- and whenever possible, we require 
already $\iota^{(2)}$ to be induced by $\iota^{(1)}$ according to (4.12).    
\sn
Having chosen such a collection of injections $\iota_{i,\,m}^{k\ (2)}$ 
for all $i,k,m \in I_q$ -- the choice involved is a finite one, to 
be discussed later -- the compatibility with concatenation also ensures 
that we can take the inductive limit of the system $\bigl(
\iota_{i,\,m}^{k\ (n)}\bigr)_n$, and we arrive at well-defined embeddings 
of infinite path spaces 
$$
\iota_{i}^{k}\,:\ \bigoplus_j \bigl( \p_{j} \bigr)^{\oplus
N_{ij}^k}   \lra  \bigl(\p_{k} \bigr)^{\oplus n_i} 
\eqno(4.13)$$ 
which by definition map elementary paths to elementary paths, preserve 
length and endpoint of every finite path and, in particular, map ground 
states to ground states. 
\sn
We need two more (canonical) maps to be able to write down a  formula for 
$C_{ij;\,\alpha}^k$. One is the projection {}from the $n_i$-fold direct 
sum of $\p_k$ onto the $\alpha$'th factor, $\alpha=1, \ldots, n_i$, 
$$
{\rm pr}_{\alpha}\,:\  \bigl(\p_{k} \bigr)^{\oplus n_i} 
\lra \p_k\ .
\eqno(4.14)$$
The other is the inclusion of $\p_j$ into the direct sum of path spaces
occurring in the fusion of $i$ and $k$ -- which, however, vanishes if 
$N_{ij}^k =0$. We write 
$$
\varepsilon_{ij}^k := \delta_{N_{ij}^k,1}\cdot\,{\rm incl}_j \,:\ 
\p_j \lra \bigoplus_{j'} \bigl( \p_{j'} \bigr)^{\oplus N_{ij'}^k} 
\eqno(4.15)$$
for this ``weighted'' inclusion. 
\mn
{\bf Proposition 4.7}\quad Denote by 
$$
\Gamma^k_{ij;\,\alpha} := {\rm pr}_{\alpha} \circ \iota^k_i \circ 
\varepsilon_{ij}^k \,:\ \p_j \lra \p_{k}  
$$
the composition of the maps (4.13-15), and define the 
$k$-$j$-strings $C_{ij;\,\alpha}^k$ by 
$$
C_{ij;\,\alpha}^k := \sum_{|p\rangle\in {\cal P}_j}  
|\,\Gamma^k_{ij;\,\alpha}(p)\, \rangle\,\langle\, p\,|  \ .
$$
These $C_{ij;\,\alpha}^k$ satisfy the assumption of Proposition 4.4.
\sn
\kapp The proof is straightforward, using that $\Gamma^k_{ij;\,\alpha}$ 
maps elementary paths to elementary paths injectively, as well as the string 
multiplication rule: 
$$\eqalignno{
\sum_{\alpha=1}^{n_i}  \bigl(C_{im;\,\alpha}^k\bigr)^* C_{ij;\,\alpha}^k
&= \sum_{\alpha=1}^{n_i}\ \sum_{|q\rangle\in {\cal P}_m}\ 
\sum_{|p\rangle\in {\cal P}_j}  |\,q\,\rangle\,\langle\,\Gamma^k_{im;\,
\alpha}(q)\,|\,\Gamma^k_{ij;\,\alpha}(p)\,\rangle\,\langle\, p\,|   
&\cr
&= \delta_{N_{ij}^k,1}  \sum_{|q\rangle\in {\cal P}_m}\ \sum_{|p\rangle\in 
{\cal P}_j} |\,q\,\rangle\,\delta_{p,q}\,\langle\, p\,| =
\delta_{N_{ij}^k,1}\,\delta_{m,j}\,\,\one_{{\cal P}_j} \ 
&\qed\cr}$$
\mn
As an aside, let us mention the following applications of this proposition, 
or of eq.\ (4.10). The operator $C_{ij}^k\,:\ \p_j \lra \p_k^{\oplus n_i}$ 
given by 
$$
C_{ij}^k=\sum_{|p\rangle\in{\cal P}_j} |\,\iota^k_i(\varepsilon_{ij}^k(p))\,
\rangle\,\langle\, p\,| \ ,
\eqno(4.16)$$
which we can also write as a column vector $C_{ij}^k =\bigl(C_{ij;\,\alpha}^k
\bigr)_{\alpha=1}^{n_i}$, is an {\sl isometry}. Furthermore, the operators
$\Pi_{ij}^k$ and $\Pi_i^k$ in $M_{n_i}(\a_k)$, defined as  
$$
\Pi_{ij}^k = C_{ij}^k \bigl(C_{ij}^k\bigr)^*\ ,\quad\quad 
\Pi_i^k = \sum_{j\in I_q} \Pi_{ij}^k\ ,
\eqno(4.17)$$ 
are both projections; in the first line, we regard $\bigl(C_{ij}^k\bigr)^*$ 
as a row vector of $j$-$k$-strings. When restricted to finite paths in 
$\p_k\n = \bigoplus_l \p_{k,l}\n$, cf.\ eq.\ (2.23), the rank of $\Pi_i^k$ 
is $\sum_j N_{ij}^k\,{\rm dim}\p_j\n\,$. These operators will become 
important later when we will discuss the amplimorphisms of the observable 
algebra.  
\mn                  Clearly, the construction 
of the strings $C_{ij;\,\alpha}^k$ we have given is not the only 
way to solve (4.10). Nevertheless, we think that {}from the point of view of 
path spaces, our procedure is the most -- and maybe even the only -- natural 
one. Besides that, requiring that the embeddings $\iota^k_i$ are compatible
with path prolongation reduces the amount of choices to be made quite 
drastically: $\iota^k_i$ is determined up to a unitary transformation in the 
finite sub-algebra $M_{n_i}(\a^{(2)}_k)$ -- with the further constraint 
that elementary paths should be mapped to elementary paths. In the example
below, it turns out that this essentially leaves only twists by certain
permutation matrices in $M_{n_i}(\C\cdot\one_{{\cal A}_k})$.
\sn
All in all, our prescription how to construct charged field multiplets in 
the amplified path field algebra leads us to an almost unique and above all
natural solution, which was possible by exploiting the ``fine structure'' of 
the underlying path spaces. 
\bn
At the end of this section, let us again take a look at the case of the 
$c(2,5)$ minimal model. There, the non-trivial field multiplet $F_1$ is an 
element of $M_2(\f)$, and according to Proposition 4.3, it has the 
following matrix structure
\def\vph{\vphantom{\sum_{y=0}^N}}
$$
\bigl((F_1)_{\alpha\beta}\bigr)_{\beta=1}^2 = 
\pmatrix{ 0\vph&C^1_{11;\,\alpha} &0&0&0&0\cr
C^1_{10;\,\alpha}\vph &0 &C^1_{11;\,\alpha}&0&0&0\cr
0\vph&0&0 &C^1_{10;\,\alpha}  &C^1_{11;\,\alpha} &0\cr}
\eqno(4.18)$$
for $\alpha=1,2$. With the help of the explicit formula (4.6) for the 
amplimorphism $\nu_1$ of ${\tt g}_{(5)}$, it is straightforward to check 
covariance (4.7) of $F_1$ explicitly -- and in this example, also uniqueness 
of the solution (4.18) is not too difficult to show. \hfbr
\noindent The strings $C_{1j;\,\alpha}^k$ are constructed as above: For 
$k=0$, we have to choose embeddings $\iota_{1,\,m}^{0\ (n)}\,:\ 
\p_{1,m}^{(n)} \lra \p_{0,m}^{(n)}\oplus\p_{0,m}^{(n)}$ for both path 
ends $m=0,1$ and for $n\geq 2$ -- or already for $n=1$ if possible: 
$\p_{0,0}^{(1)} = \emptyset$, but $\iota_{1,\,1}^{0\ (1)}$ can be defined: 
It maps the edge $(1\rightarrow 1)\in\p_{1,1}^{(1)}$ to one of the two 
copies of $(0\rightarrow 1)$ present in $\p_{0,1}^{(1)}\oplus
\p_{0,1}^{(1)}\,$. This determines $\iota_{1}^{0\ (n)}$ on all paths in 
$\p_1^{(n)}$ that go through the node 1 on the first floor. In particular, 
one of the two copies of the path $(0;1,1)$ in $\p_{0,1}^{(2)}\oplus
\p_{0,1}^{(2)}$ is in the image of those paths, so we have to map the 
remaining path $(1; 0, 1)\in\p_{1,1}^{(2)}$ to the other copy of $(0;1,1)$ 
in $\p_{0,1}^{(2)}\oplus\p_{0,1}^{(2)}$. This defines 
$\iota_{1,\,m}^{0\ (2)}$ and therefore $\iota^0_1$ completely.  \hfbr 
\noindent For $k=1$, we need to define 
$\iota_1^1\,:\ \p_0 \oplus \p_1 \lra \p_1 \oplus\p_1$. 
The only natural possibility is to identify the space $\p_1$ on the
left with one of the $\p_1$'s on the right, and then map $\p_0$ into 
the second copy of $\p_1$
with the ``initial condition'' $\iota_{1,\,1}^{1\ (1)}\,(0\rightarrow1)= 
(1\rightarrow1) \in \p_{1,1}^{(1)}$. Then, the only choice  in the
construction is  which copy of $\p_1$ on the right to identify with  
the $\p_1$ on the left. \hfbr 
\noindent   This means that both embeddings $\iota_1^k$ are determined 
up to a permutation  matrix in $M_2(\C)$, acting trivially within the 
path representation spaces $\p_k$.  Strictly speaking, however, we are 
not forced to map the space $\p_1$ in $\p_0\oplus\p_1$ ``as a whole''into 
one of the $\p_1$ in the target space, but we could also ``distribute'' it 
over both copies. This relatively unnatural choice would introduce a higher
degree of indeterminacy into our construction. Note, however, that in any 
case we can arrange $\iota_1^k$ so as to map lowest energy states 
(the sequences of nodes  $(0; 1, 1,1\ldots)$ or $(1; 1, 1,1\ldots)\,$) to 
lowest energy states again. 
\bn\bn
\vfil\eject
\leftline{\bf 5. Global amplimorphisms}
\bn
It is now very esay to write down global amplimorphisms of the path 
observable algebra which implement the charged sectors of our models. 
Copying the procedure of \q{62}, we associate to each field multiplet 
$F_i \in M_{n_i}(\f)$ the linear map 
$$
\rho_i\,:\ \cases{&$\a \lra M_{n_i}(\f)$ \cr \nvs
           &$A \longmapsto \rho_i(A) = 
     \Bigl(\rho_i(A)_{\alpha\beta}\Bigr)_{\alpha,\beta=1}^{n_i} $\cr}
\eqno({\rm 5.1a})$$
with 
$$
\rho_i(A)_{\alpha\beta} := \sum_{\gamma=1}^{n_i} \bigl(F_i\bigr)_{\alpha
\gamma}\, A\, \bigl( F_i^*\bigr)_{\gamma\beta}\ .
\eqno({\rm 5.1b})$$
Here and in the following, \a\ is viewed as sub-algebra of \f\ by the 
diagonal embedding as in the proof of Proposition 3.7. 
We list the relevant properties of these maps in a series of propositions. 
First we have to show that the maps $\rho_i$ deserve the 
name {\sl amplimorphisms of the global observable algebra}:  
\mn
{\bf Proposition 5.1}\quad The map $\rho_i$ takes values in the $n_i\,$th 
amplification of the global path observable algebras \a. It  
is an injective ${}^*$-homomorphism of AF-algebras. 
\mn
\kapp By definition, $\rho_i(A)_{\alpha\beta}$ is an element of \f\ for all 
$A\in\a$. In order to show that $\rho_i(A)_{\alpha\beta}$ are observables, 
it is sufficient to check that they commute with $U(a)$ for all $a\in 
{\tt g}$, see Proposition 3.7. Identifying $a$ with $U(a)$ for simplicity, 
and using the summation convention, we compute for arbitrary $A\in\a$   
$$\eqalign{
a\cdot \rho_i(A)_{\alpha\beta} & =a (F_i)_{\alpha\gamma}\,A\,(F_i^*)_{
\gamma\beta} 
= (F_i)_{\alpha\delta} (\nu_i(a))_{\delta\gamma}\, A\, (F_i^*)_{\gamma\beta}
= (F_i)_{\alpha\delta}\,A\,(\nu_i(a))_{\delta\gamma}  (F_i^*)_{\gamma\beta}
\cr
&=(F_i)_{\alpha\delta}\,A\, \bigl(F_i \nu_i(a)^*\bigr)^*_{\delta\beta}
= (F_i)_{\alpha\delta}\,A\, \bigl(a^* F_i\bigr)^*_{\delta\beta}
= (F_i)_{\alpha\delta}\,A\, (F_i^*)_{\delta\beta}\, a 
\cr
&= \rho_i(A)_{\alpha\beta} \cdot a\ ;
\cr}$$
we have used Proposition 3.7 and the properties (4.7) and 
(4.8) of the covariant field multiplets repeatedly. \hfbr
\noindent By construction, it is clear that $\rho_i$ respects the 
${}^*$-operations of the path algebras,  
and multiplicativity is again established with the 
help of the covariance and completeness relations of the field multiplets: 
$$\eqalign{
\rho_i(A)_{\alpha\gamma}\rho_i(B)_{\gamma\beta} &= 
(F_i)_{\alpha\delta}\,A\,(F_i^*)_{\delta\gamma}\,
(F_i)_{\gamma\epsilon}\,B\,(F_i^*)_{\epsilon\beta}
\cr
&= \bigl( F_i\cdot
\nu_i(1)\bigr)_{\alpha\epsilon}\,AB\,(F_i^*)_{\epsilon\beta}
= \bigl(\rho_i(AB)\bigr)_{\alpha\beta} 
\cr}$$
for all elements $A, B$ of the path observable algebra $\a\subset \f$.  \hfbr
\noindent That $\rho_i$ is an injective map {}from \a\ to $M_{n_i}(\a)$ also 
follows directly {}from (4.8): ``Sandwiching'' $\rho_i(A)$ 
by $F_i^*$ and $F_i$, and using the multiplicativity of the 
{\tt g}-amplimorphism $\nu_i$, we obtain 
$$
\bigl( F_i^*\,\rho_i(A)\,F_i\bigr)_{\alpha\beta} = 
\nu_i(1)_{\alpha\gamma}\,A\,\nu_i(1)_{\gamma\beta} = A\,\nu_i(1)_{\alpha\beta}
$$
for all $\alpha, \beta=1,\ldots, n_i$. Thus, $\rho_i(A)=0$ 
implies $A=0$. \hfill\qed
\mn
Note that these properties do not hold if we extend $\rho_i$ to 
all of \f\ -- and also that the proof did not depend on any specific 
properties tied to the path 
representations. Those came into play when we had to solve the conditions 
(4.7) and (4.8) for the field multiplets explicitly by constructing 
the strings $C_{ij;\,\alpha}^k\,$, which appear in the concrete 
expressions for the amplimorphisms. 
\mn
In order to obtain such formulas, we use 
the decomposition $A=A_0+\ldots+A_K$ of elements of $\a=\a_1\oplus\ldots
\oplus\a_K$, and the diagonal embedding (3.15) of \a\ into the field algebra. 
Inserting the covariant field multiplets {}from Proposition 4.3 into (5.1), 
and keeping track of the Kronecker symbols, 
yields the following expressions for the elements of the amplimorphisms: 
$$
\bigl(\rho_i(A)\bigr)_{(\alpha; n_k,\sigma)(\beta; n_j,\tau)} 
= \delta_{k,j}\,\delta_{\sigma,\tau}\,
\sum_{l\in I_q} \delta_{N_{il}^k,1}\, 
C_{il;\,\alpha}^k\,A_l\,\bigl(C_{il;\,\beta}^k\bigr)^*
\eqno(5.2)$$
with $C_{il;\,\alpha}^k$ as in Proposition 4.7; the matrix notation is as 
in section 4.1. By construction of 
the string coefficients, $\rho_i$ maps the $n$-floor finite-dimensional 
sub-algebra $\a\n$ of \a\ into $M_{n_i}(\a\n)$ and is compatible with 
the embeddings $\Phi\n$ induced by path concatenation. We say that the 
amplimorphisms {\sl preserve the filtration} of the systems 
$\bigl(\a_i\n, \Phi\n\bigr)$. 
\sn 
In addition, formula (5.2) makes it clear that the amplimorphisms 
$\rho_i$ are {\sl non-unital} in general. We have 
$$
\rho_i(\one)_{(\alpha; n_k,\sigma)(\beta; n_j,\tau)} 
= \delta_{k,j}\,\delta_{\sigma,\tau}\, \bigl( \Pi_i^k\bigr)_{\alpha\beta} 
$$ 
with the projection $\Pi_i^k \in M_{n_i}(\a_k)$ {}from eq.\ (4.17), which is 
$\neq\one$ for $i\neq0$. 
\mn
During the construction of the field multiplets, we have made several 
choices, and we must determine to which extend the amplimorphisms $\rho_i$ 
depend on them: 
\sn
{\bf Proposition 5.2}\quad The \a-amplimorphisms $\rho_i$ depend only on 
the isomorphism class of the {\tt g}-amplimorphisms $\nu_i$ which enter 
the covariance law (4.7) of the field multiplets. For a given class of 
$\nu_i$, $\rho_i$ are unique up to conjugation by unitaries in $M_{n_i}(\a)$. 
\sn
\kapp The second statement follows immediately {}from the $C_{ij;\,\alpha}^k\,$
dependence of the \a-amplimorphisms. 
To see the first claim, assume that we ``twist'' the {\tt g}-amplimorphism 
$\nu_i$ by a unitary $u\in M_{n_i}({\tt g})$ and use 
$\nu_i' := {\rm Ad}_u\circ \nu_i$ instead. 
Then the covariant field multiplets change as 
$F_i' := F_i\cdot u^{-1}$, where $u^{-1}$ acts through the representation 
$U$ of {\tt g} on $\h^{\rm tot}$, see Definition 3.6. Since $U(u)$ 
commutes with all elements of \a, formula (5.1) for $\rho_i$ stays 
invariant under a twist.   \hfill\qed
\mn
The isomorphism classes of the {\tt g}-amplimorphisms were of course 
fixed by the \ctq\ fusion rules {}from the start, therefore we conclude 
that our construction yields filtration-preserving amplimorphisms of 
the global path observable algebra of unique isomorphism type. But moreover, 
in view of the remarks on the choice of strings $C_{ij;\,\alpha}^k\,$ made 
after Proposition 4.7, the $\rho_i$ which are ``natural'' {}from the 
path perspective are fixed up to conjugation by a permutation matrix 
in $M_{n_i}(\a^{(2)})$.  
\mn
We can now ask whether the $\rho_i$ implement the representations of the 
global observable algebra on the vacuum sector:
\sn 
{\bf Proposition 5.3}\quad Let $\pi_i\,:\ \a\lra\a_i$ be the 
projection of \a\ onto the representation in the sector $\h_i$, and denote 
the extensions to amplifications $M_n(\a)$ by the same symbol. 
Then the amplimorphisms $\rho_i$ of eq.\ (5.1) satisfy 
$$
\pi_0\circ\rho_i  \simeq \pi_i\ , 
$$
where $\simeq$ denotes equivalence by  isometries. 
\sn
\kapp The statement follows immediately {}from formula (5.2). Setting 
$k=0$, the condition $N_{il}^0=1$ enforces $l=i$ (uniqueness of 
the conjugated sector), and we remain with 
$$
\pi_0\bigl(\rho_i(A)\bigr)_{\alpha\beta} = 
C_{ii;\,\alpha}^0\,A_i\,\bigl(C_{ii;\,\beta}^0\bigr)^*\ .
$$
Thus, the string isometries $C_{ii}^0$ {}from eq.\ (4.16)  `transport'' 
the representation $\a_i$ into $M_{n_i}(\a_0)$ and 
implement the equivalence.  Note that by our construction of the 
$C_{ij;\,\alpha}^k$ the equivalence also holds when \a\ and 
$\a_i$ are replaced by finite-dimensional sub-algebras $\a\n$ and 
$\a_i\n$.                      \hfill\qed 
\mn
{\bf Proposition 5.4}\quad Upon composition, the amplimorphisms realize 
the fusion rules of the \ctq\ minimal models, i.e.\ the relations  
$$
\pi_0\circ\,(\rho_i\circ\rho_j)\ \  \simeq\ \  
\bigoplus_{k\in I_q}\, \bigoplus_{r=1}^{N_{ij}^k}\  \pi_0 \circ \rho_k 
$$ 
hold in $M_{n_in_j}(\a_0)\,$. 
\sn
\kapp We apply the lhs to $A\in\a$ and find after a short calculation  
$$
\pi_0\bigl(\rho_i(\rho_j(A))\bigr)_{\alpha\alpha',\beta\beta'} 
= \sum_{k\in I_q} \delta_{N_{ij}^k,1} \,    
\bigl( C_{ii;\,\alpha}^0\,C_{jk;\,\alpha'}^i \bigr) \,A_k\,
\bigl( C_{ii;\,\beta}^0\,C_{jk;\,\beta'}^i \bigr)^* 
$$
where $\alpha, \beta=1,\ldots, n_i$ and $\alpha', \beta'=1,\ldots,n_j$. 
Again, the claim follows {}from the completeness and orthogonality 
relation (4.10) for the strings $C_{ij;\,\alpha}^k$.  \hfill\qed 
\mn
In view of all these properties, we may regard the amplimorphisms 
$\rho_i$ as variants of the DHR morphisms of algebraic quantum 
field theory. 
\bn\bn
\vfil\eject
\leftline{\bf 6. Open problems}
\bn
Having constructed amplimorphisms for the \ctq\ models, one may of 
course ask whether they yield further information on these conformal 
field theories. Since the algebraic approach is superior to all others 
when it comes to the discussion of braid group statistics, one should 
in particular try to compute intertwiners (statistics operators), 
left inverses and Markov traces associated to the morphisms of 
the observable algebra. In this way, interesting braid group 
representations and knot invariants might arise and, as a by-product, one 
could supply the QSA {\tt g} with the full data of a weak quasi-triangular 
quasi Hopf algebra. The results of \q{62} indicate that dealing with 
non-unital amplimorphisms rather that unital endomorphisms does not pose 
severe problems, and we hope that this program can be carried out even in 
the absence of local information in our path algebras and morphisms. 
\sn
Nevertheless, if we aim at a complete description of the \ctq\ models 
within the algebraic framework, it is important to recover 
the local net structure inside the global algebras, and to show that 
our amplimorphisms are equivalent to localizable ones. One possible 
starting point for the construction of local sub-algebras of the path algebra 
is provided by the su(1,1) action on the path spaces constructed in \q{56}. 
But although there seems to be no principle problem, 
it is technically rather difficult to implement the conditions of M\"obius
covariance on the local sub-algebras. \hfbr
\noindent In this context, we conjecture that it is precisely because they 
are filtration-preserving that our global amplimorphisms have good chances 
to be equivalent to covariant and localized morphisms: The su(1,1)-action 
of \q{56} is designed in analogy with the excitation of quasi-particles 
and therefore respects the finite length filtration of the path spaces as 
much as possible.   \hfbr
\noindent Given an action of su(1,1) or even the Virasoro algebra 
on the path spaces, we could also make our somewhat abstract amplimorphisms 
more concrete: Although their action on path algebra elements can be 
made as explicit as we wish, it would be interesting to have formulas 
for $\rho_i$ applied to Virasoro generators, similar to \q{44} where 
neat expressions could be obtained because the Virasoro modes 
have simple expansions in terms of the fermion modes, which in turn 
are acted on by the endomorphisms in a simple fashion. 
\sn
In addition, once sub-algebras of local observable have been identified 
within the global path algebra, one would also like to make contact to 
the von Neumann algebra description of local QFTs with all its particular 
merits. It remains to be seen whether one then meets problems with the 
non-unitarity of the \ctq\ models, which apparently played no role 
at the purely algebraic level of our global considerations. 
\mn
We have seen in this paper that the path representations of the \ctq\ 
minimal models open up a lot of interesting possibilities. They allow 
to organize a great deal of information on these CFTs into a single 
labeled graph; they naturally lead to an AF-algebraic description 
of theories which a priori are defined in terms of unbounded Virasoro 
modes; they can, in this respect, be regarded as an alternative to the 
usual free field constructions; they seem to encode, 
via the quasi-particle reformulation, structural details of  
non-conformal relatives of the \ctq\ minimal models within the CFT 
and might, therefore, even be useful in the context of massive 
integrable quantum field theories.  \hfbr
\noindent    In view of these facts, it is desirable to find similar 
path representations for other conformal models 
as well. In \q{41}, this has been achieved for a subset of modules in the 
$c({\rm even}, {\rm odd})$ Virasoro minimal models, with our graphs $\g_q$ 
again playing an important role. However, since some of the sectors do not 
have a path description, our construction cannot be applied to those
models, yet. 
\sn
Whereas the results in \q{41} have been obtained by factorization of some 
of the characters of the minimal models, i.e.\ by purely combinatorial 
means, one could alternatively try to imitate the FNO procedure in order 
to determine explicit bases of the irreducible modules. However, for  
general minimal models the structure of the annihilating ideal is more 
involved than in the \ctq\ cases, and it seems necessary to pass to 
the maximally extended chiral observable algebra before progress can 
be made. 
\mn
We also came across some problems of a more abstract nature: In the
introduction, we raised the speculation that there is a general relation 
between AF-algebras and conformal field theories. Somewhat related to 
this conjecture, we have seen that the global path observable algebras 
of the \ctq\ models are of the same type as the string algebras which show 
up as intertwiner (symmetry) algebras in the DHR framework. Thus it seems 
that in the case of these conformal models, internal and space-time 
symmetries are indeed ``inexorably linked'' \q{59}. Still, the relationship 
remains to be made precise. On the other hand, we have shown that there also 
is a canonical semi-simple QSA for any quasi-rational CFT. The question 
is whether one can  find an axiomatic foundation 
of the notions introduced in \q{48}. Since the dimensions of the 
small spaces provide new invariants of quasi-rational CFTs, this might lead 
to interesting developments within algebraic QFT and even in the 
theory of operator algebras. 
\bn\vskip 2.5cm\bn
{\bf Acknowledgements} 
\bn
I would like to thank A.\ Alekseev, D.\ Buchholz, K.\ Fredenhagen, 
J.\ Fr\"ohlich, J.\ Fuchs, A.\ Ganchev, K.\ Gaw\c edzki, W.\ Nahm, 
V.\ Schomerus, B.\ Schroer, K.\ Szlachanyi and P.\ Vecsernyes for 
useful discussions, comments and encouragement. 
I am particularly indebted to M.\ R\"osgen whose collaboration 
on path representations was invaluable.   \hfbr
\noindent This work was partially support by a HCM Fellowship 
of the European Union.
\vfil\eject
\vbox{
\noindent{\bf Figure 1: }Fusion graphs $\g_q$ of the minimal dimension field 
for $q=5,\ 7$ and 9; the labels $i$ at the nodes refer to the 
sectors $\phi_i\,$: 
\mn
\hbox{\phantom{XXX}\epsfbox{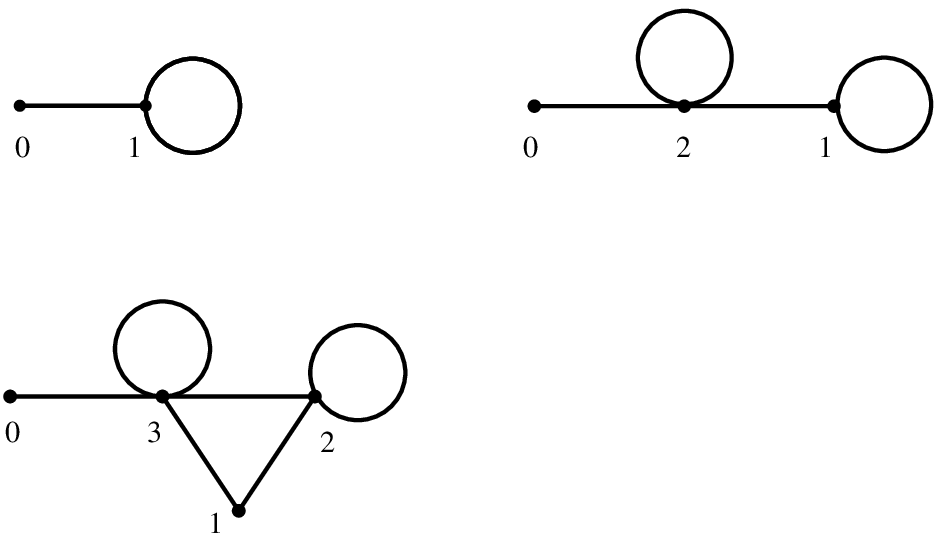}} 
\bn\bn\bn
{\bf Figure 2: }Bratteli diagrams $\b_{5,0}$, $\b_{5,1}$ and 
$\widehat{\b}_5$, associated to the observable algebras in the vacuum 
sector, in the $h=-{1\over5}$ representation, and to the path field algebra 
of the \break 
\sn\vskip-15pt
{\rightskip=140pt \noindent $c(2,5)$ model. Labels on the $n\,$th
floor give the dimensions of the spaces 
of paths running in $n$ steps {}from the top node (on the zeroth or $-2\,$nd
floor of the diagram) to the  node 0 (on the left) or 1 (on the right) of
$\g_5\,$:\par}
\vskip-1.2cm
\hbox{\phantom{}\epsfbox{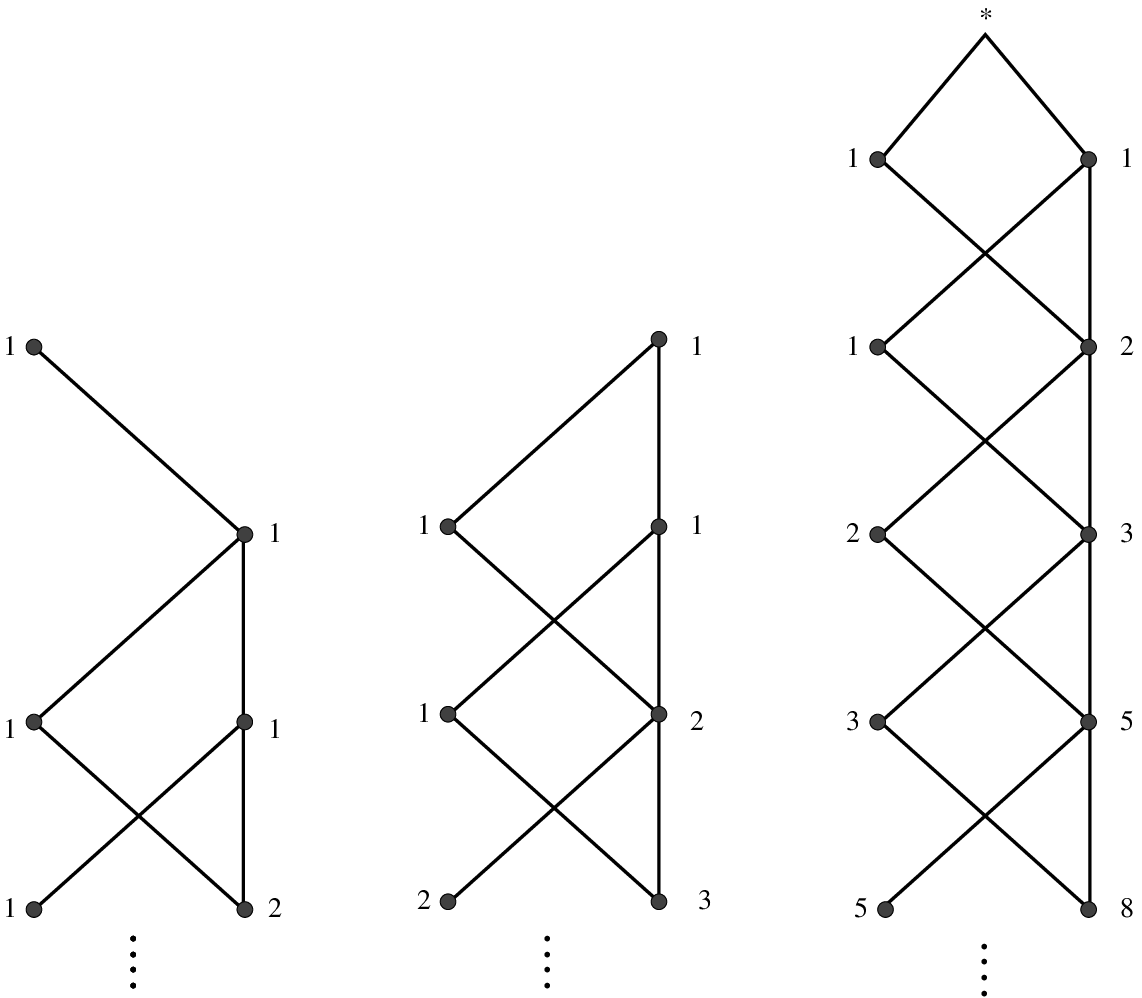}} 
}
\eject
{\klein  

\def\CMP#1{{\klein Commun.\ Math.\ Phys.\ {\bfk #1}}}
\def\intj{{\klein Int.\ J.\ Mod.\ Phys.\ A\ }}
\def\np{{\klein Nucl.\ Phys.\ B\ }}
\def\pl{{\klein Phys.\ Lett.\ B\ }}

\parindent=13pt \def\sn{\smallskip\vskip-2pt\noindent}
\leftline{{\bf References}}
\sn 
{\baselineskip=10pt {\klein 
\def\q#1{\item{{\klein #1.}}}
\q{1}A.Yu.\ Alekseev, L.D.\ Faddeev, M.A.\ Semenov-Tian-Shansky,
   {\itk Hidden quantum groups inside Kac-Moody algebras},
     \CMP{149} (1992) 335-345 \sn
\q{2}A.Yu.\ Alekseev, A.\ Recknagel, {\itk The embedding structure and 
   the shift operator of the U(1) lattice current algebra}, Lett.\ Math.\ 
   Phys.\ {\bfk37} (1996) 15-27 \sn
\q{3}A.Yu.\ Alekseev, A.\ Recknagel, V.\ Schomerus, {\itk Generalization of 
   the Knizhnik-Zamo\-lodchikov equations}, preprint, 13 pp., 
   hep-th/9610066 \sn 
\q{4}L.\ Alvarez-Gaum\'e, C.\ Gomez, G.\ Sierra, {\itk Quantum 
   group interpretation of some conformal field theories}, 
   Phys.\ Lett.\ B {\bfk220} (1989) 142-152 \sn
\q{5}G.E.\ Andrews, {\itk q-series: their development and application
   in analysis, number theory, combinatorics, physics, and computer
   algebra}, Conf.\ Board of the Math.\ Sciences, 
   Reg.\ Conf.\ Ser.\ in Math.\ {\bfk66} (1986) \sn 
\q{6}G.E.\ Andrews, R.J.\ Baxter, P.J.\ Forrester, {\itk Eight-vertex 
  SOS model and generalized Rogers-Ramanujan-type identities},  
   J.\ Stat.\ Phys.\ {\bfk35} (1984) 193-266 \sn
\q{7}M.\ Bauer, P.\ Di Francesco, C.\ Itzykson, J.-B.\ Zuber, {\itk 
   Covariant differential equations and singular vectors in Virasoro 
   representations},  Nucl.\ Phys.\ B {\bfk362} (1991) 515-562\sn
\q{8}A.A.\ Belavin, A.M.\ Polyakov, A.B.\ Zamolodchikov,
   {\itk Infinite conformal symmetry in two-di\-men\-sio\-nal
   quantum field theory}, \np {\bfk 241} (1984) 333-380 \sn
\q{9}L.\ Benoit, Y.\ Saint-Aubin, {\itk Degenerate conformal
   field theories and explicit expressions of some null 
   vectors}, Phys.\ Lett.\ B {\bfk215} (1988) 517-522 \sn
\q{10}A.\ Berkovich, B.M.\ McCoy, A.\ Schilling, {\itk Rogers-Schur-Ramanujan 
   type identities for the ${\scriptstyle M(p,p')}$ minimal models of 
   conformal field theory}, Bonn and Stony Brook preprint, 73 pp., 
   q-alg/9607020\sn
\q{11}B.\ Blackadar, {\itk K-Theory for operator algebras}, MSRI
   Publ.\ 5, New York 1986 \sn
\q{12}R.\ Blumenhagen, M.\ Flohr, A.\ Kliem, W.\ Nahm, A.\ Recknagel, 
   R.\ Varnhagen, {\itk W-algebras with two and three generators},
   \np{\bfk 361} (1991) 255-289 \sn
\q{13}J.\ B\"ockenhauer, {\itk Localized endomorphisms of the 
   chiral Ising model}, \CMP{177} (1996) 265-304 \sn 
\q{14}J.\ B\"ockenhauer, J.\ Fuchs, {\itk Higher level WZW sectors {}from 
   free fermions}, DESY preprint, 31 pp., hep-th/9602116\sn
\q{15}D.\ Buchholz, G.\ Mack, I.T.\ Todorov, {\itk The current
   algebra on the circle as a germ of local field theories},
   \np (Proc.\ Suppl.) {\bfk5B} (1988) 20-56 \sn
\q{16}J.L.\ Cardy, {\itk Conformal invariance and the Yang-Lee edge 
   singularity  two dimensions}, Phys.\ Rev.\ Lett.\ {\bfk54} (1985) 
   1354-1356 \sn 
\q{17}M.\ Caselle, G.\ Ponzano, F.\ Ravanini,
   {\itk Towards a classification of fusion rule algebras in rational 
   conformal field theories}, Int.\ J.\ Mod.\ Phys.\ B  {\bfk 6} (1992) 
   2075-2090 \sn  
\q{18}S.\ Doplicher, R.\ Haag, J.E.\ Roberts, {\itk Local
   observables  and particle statistics I,II}, \CMP{23} (1971) 199-230,
    {\bfk35} (1974) 49-85 \sn
\q{19}S.\ Doplicher, J.E.\ Roberts, {\itk A new duality theory for 
   compact groups}, Invent.\ Math.\ {\bfk98} (1989) 157-218; \ \ 
   {\itk Why there is a field algebra with a compact gauge group 
   describing the superselection structure in particle physics}, 
   Commun.\ Math.\ Phys.\ {\bfk 131} (1990) 51-107\sn 
\q{20} V.G.\ Drinfeld, {\itk Quasi-Hopf algebras and 
    Knizhnik-Zamolodchikov equations}, in:\ {\itk Problems of modern 
    quantum field theory}, Proceedings Alushta 1989, Research reports 
    in physics, Springer 1989; {\itk Quasi-Hopf algebras}, 
    Lenin\-grad Math.\ J.\ Vol.\ {\bfk 1} (1990) No.\ 6 \sn 
\q{21}B.L.\ Feigin, D.B.\ Fuchs,
   {\itk Invariant skew-symmetric differential operators on
   the line and Verma modules over the Virasoro algebra},
   Funct.\ Anal.\ Appl.\ {\bfk 16} (1982) 114-126;\ \   {\itk Verma 
   modules over the Virasoro algebra}, Lect.\ Notes in
   Math.\ {\bfk 1060}, 230-245, Springer 1984 \sn
\q{22}B.L.\ Feigin, T.\ Nakanishi, H.\ Ooguri, {\itk The annihilating ideals 
  of minimal models}, \intj {\bfk7} Suppl.\ {\bfk 1A} (1992) 217-238 \sn
\q{23}G.\ Felder, J.\ Fr\"ohlich, G.\ Keller, {\itk On the structure of 
  unitary conformal field theory I,II}, Commun.\ Math.\ Phys.\ 
  {\bfk124} (1989) 417-463, {\bfk130} (1990) 1-49 \sn 
\q{24}K.\ Fredenhagen, K.-H.\ Rehren, B.\ Schroer,
  {\itk Superselection sectors with braid group statistics and exchange 
  algebras I,II}, \CMP{125} (1989) 201-226, Rev.\ Math.\ Phys.\ Special 
  issue (1992) 111-154\sn
\q{25}P.G.O.\ Freund, T.R.\ Klassen, E.\ Melzer, {\itk S-Matrices for 
   perturbations of certain conformal field theories}, \pl {\bfk229} 
   (1989) 243-247 \sn
\q{26}D.\ Friedan, Z.\ Qiu, S.H.\ Shenker, {\itk Conformal invariance, 
   unitarity and two-dimensional critical exponents}, 
   Phys.\ Rev.\ Lett.\ {\bfk 52} (1984) 1575-1578 \sn
\q{27}J.\ Fr\"ohlich, {\itk Statistics of fields, the Yang-Baxter
  equation and the theory of knots and links}, in:\  {\itk
  Non-perturbative quantum field theory}, eds.\ G.\ t'Hooft,
  A.\ Jaffe, G.\ Mack, P.K.\ Mitter and R.\ Stora, Plenum 1988 \sn
\q{28}J.\ Fr\"ohlich, F.\ Gabbiani, {\itk Braid statistics in local quantum 
  theory}, Rev.\ Math.\ Phys.\ {\bfk2} (1990) 251-353\sn
\q{29}J.\ Fr\"ohlich, T.\ Kerler, {\itk Quantum groups, quantum categories 
   and quantum field theory}, Lect.\ Notes in Math.\ {\bfk1542}, Springer 
   1993\sn
\q{30}J.\ Fuchs, {\itk Algebraic conformal field theory},
   Proceedings Gosen 1991, Ahrenshoop Symp.\ 1991, 99-114 \sn
\q{31}J.\ Fuchs, A.\ Ganchev, P.\ Vecsernyes, {\itk Level 1 WZW superselection
   sectors}, Commun.\ Math.\ Phys.\ {\bfk146} (1992) 553-584;\ \ {\itk Simple 
   WZW superselection sectors}, Lett.\ Math.\ Phys.\ {\bfk28} (1993) 31-42\sn
\q{32}J.\ Fuchs, A.\ Ganchev, P.\ Vecsernyes, {\itk Towards a classification 
   of rational Hopf algebras}, NIKHEF preprint, 44 pp.,  hep-th/9402153; \ \ 
   {\itk On the quantum symmetry of rational field theories}, Theor.\ 
   Math.\ Phys.\ {\bfk98} (1994) 266-276 \sn 
\q{33}F.\ Gabbiani, J.\ Fr\"ohlich, {\itk Operator algebras and 
  conformal field theory}, Commun.\ Math.\ Phys.\ {\bfk155} (1993) 
  569-640\sn
\q{34}M.\ Gaberdiel, {\itk Fusion of twisted representations}, Cambridge 
   preprint, 23 pp., hep-th/9607036 \sn
\q{35}F.M.\ Goodman, P.\ de la Harpe, V.F.R.\ Jones, {\itk Coxeter
   graphs and towers of algebras}, MSRI Publ.\ 14, New York 1989 \sn
\q{36}R.\ Haag, {\itk Local quantum physics}, Springer 1992\sn
\q{37}R.\ Haag, D.\ Kastler, {\itk An algebraic approach to
   field theory}, J.\ Math.\ Phys.\ {\bfk5} (1964) 848-861\sn
\q{38}R.\ Kedem, T.R.\ Klassen, B.M.\ McCoy, E.\ Melzer,
   {\itk Fermionic quasiparticle representations
   for characters of ${\scriptstyle G^{(1)}_1 \times G^{(1)}_1/G^{(1)}_2}$},
   \pl {\bfk304} (1993) 263-270;\ \  {\itk Fermionic sum representations   
   for conformal field theory characters}, \pl {\bfk307} (1993) 68-76 \sn
\q{39}R.\ Kedem, B.M.\ McCoy,
   {\itk Construction of modular branching functions {}from Bethe's 
   equations in the 3-state Potts chain}, Stony Brook preprint,  
   34 pp., hep-th/9210129 \sn
\q{40}J.\ Kellendonk, A.\ Recknagel, {\itk Virasoro representations
        on fusion graphs}, \pl {\bfk298} (1993) 329-334 \sn
\q{41}J.\ Kellendonk, M.\ R\"osgen, R.\ Varnhagen,  {\itk Path
    spaces and W-fusion in minimal models}, Int.\ J.\ Mod.\ 
    Phys.\ A {\bfk9} (1994) 1009-1024 \sn
\q{42}V.G.\ Knizhnik, A.B.\ Zamolodchikov, {\itk Current
   algebra and Wess-Zumino model in two dimensions}, Nucl.\ 
 Phys.\ B {\bfk247} (1984) 83-103\sn
\q{43}R.\ Longo, {\itk Index for subfactors and statistics of quantum
  fields I,II}, \CMP{126} (1989) 217-247, {\bfk130} (1990) 285-309 \sn
\q{44}G.\ Mack, V.\ Schomerus, {\itk Conformal field algebras with
   quantum symmetry {}from the theory of super\-selection sectors},
    \CMP{134} (1990) 139-196 \sn
\q{45}G.\ Mack, V.\ Schomerus, {\itk Quasi-Hopf quantum symmetry  
     in quantum theory}, Nucl.\ Phys.\ B {\bfk370} (1991) 185-230\sn
\q{46}G.\ Moore, N.\ Seiberg, {\itk Polynomial equations for
   rational conformal field theories}, \pl {\bfk 212} (1988) 451-460;\ \ 
   {\itk Classical and conformal quantum field theory}, 
   Commun.\ Math.\ Phys.\ {\bfk123} (1989) 177-254 \sn
\q{47}W.\ Nahm, {\itk Chiral algebras of two-dimensional chiral field theories
    and their normal ordered products}, in:\ {\itk Recent developments in
    conformal field theories}, Proc.\ Trieste October 1989, Proc.\ 3 Reg.
    Conf.\ on Math.\ Phys., Islamabad 1989 \sn
\q{48}W.\ Nahm, {\itk Quasi-rational fusion products}, Int.\ J.\ 
    Mod.\ Phys.\ B {\bfk8} (1994) 3693-3702  \sn
\q{49}A.\ Ocneanu, {\itk Quantized groups, string algebras and Galois
   theory for algebras}, in:\ {\itk Operator algebras and applications II}, 
   London Math.\ Soc.\ vol.\ 135, Cambridge 1988; \ \  {\itk Quantum symmetry, 
   differential geometry of finite graphs and classification of subfactors}, 
   University of Tokyo Seminary Notes 45, recorded by Y.\ Kawahigashi,
   July 1990\sn 
\q{50}V.\ Pasquier, {\itk Two-dimensional critical systems labeled by 
   Dynkin diagrams}, \np {\bfk285} (1987) 162-172; \ \ {\itk Etiology of 
   IRF models}, \CMP{118} (1988) 355-364\sn
\q{51}V.\ Pasquier, H.\ Saleur, {\itk Common structures between finite
   systems and conformal field theories through quantum groups},
   \np {\bfk330} (1990) 523-556 \sn
\q{52}K.-H.\ Rehren, {\itk Markov traces as characters for local
   algebras}, \np (Proc.\ Suppl.) {\bfk 18B} (1990) 259-268; \ \  {\itk Braid
   group statistics and their superselection rules}, in:\ {\itk The algebraic
   theory of superselection sectors.\ Introduction and recent results}, ed.\ 
   D.\ Kastler, World Scientific 1990\sn
\q{53}K.-H.\ Rehren, {\itk Field operators for anyons and plektons},
    \CMP{145} (1992) 123-148 \sn
\q{54}K.-H.\ Rehren, {\itk Quantum symmetry associated with braid group
   statistics}, in:\ {\itk Quantum groups}, Proc.\ Clausthal 1989, eds.\ 
   H.-D.\ Doebner et al., Lect.\ Notes in Phys.\ {\bfk370}, 318-339, 
   Springer 1990; \ \  {\itk Quantum symmetry associated with braid group
   statistics II}, in:\ {\itk Quantum symmetries}, Proc.\ Clausthal 1991, 
   eds.\ H.-D.\ Doebner et al., 14-23, World Scientific 1993 \sn 
\q{55}K.-H.\ Rehren, {\itk Subfactors and coset models}, in:\ {\itk 
   Generalized symmetries in physics}, Proc.\ Claus\-thal 1993, 
   hep-th/9308145;\ \  {\itk On the range of the index of subfactors}, 
   Vienna preprint ESI-93-14, 9 pp. \sn
\q{56}M.\ R\"osgen, {\itk Pfaddarstellungen minimaler Modelle},
   Diplomarbeit BONN-IR-93-24; \hfill
\item{}M.\ R\"osgen, R.\ Varnhagen, {\itk Steps towards lattice 
   Virasoro algebras: su(1,1)}, \pl {\bfk350} (1995) 203-211 \sn 
\q{57}A.N.\ Schellekens, S.\ Yankielowicz,
    {\itk Extended chiral algebras and modular invariant
    partition functions},     \np{\bfk327} (1989) 673-703 \sn
\q{58}V.\ Schomerus, {\itk Construction of field algebras 
   with quantum symmetry {}from local observables}, Commun.\ Math.\ 
   Phys.\ {\bfk 169} (1995) 193-236 \sn 
\q{59}B.\ Schroer, {\itk Reminiscences about many pitfalls and some 
   successes of QFT within the last three decades}, Rev.\ Math.\ Phys.\
   {\bfk7} (1995) 645-688\sn
\q{60}K.\ Szlachanyi, P.\ Vecsernyes, {\itk Quantum symmetry and braid
   group statistics in G-spin models}, \CMP{156} (1993) 127-168 \sn  
\q{61}M.\ Terhoeven, {\itk Lift of dilogarithm to partition
   identities}, Bonn preprint, 8 pp., hep-th/9211120 \sn
\q{62}P.\ Vecsernyes, {\itk On the quantum symmetry of the chiral Ising 
  model}, \np{\bfk 415} (1994) 557-588 \sn
\q{63}A.\ Wassermann, {\itk Operator algebras and conformal field theory}, 
  Proceedings ICM Z\"urich 1994, Birk\-h\"auser;\ \ {\itk Operator algebras 
  and conformal field theory II,III}, Cambridge preprints\sn
\q{64}H.W.\ Wiesbrock, {\itk Conformal quantum field theory and half sided 
  modular inclusions of von Neumann algebras}, Commun.\ Math.\ Phys.\ 
  {\bfk158} (1993) 537-544;\ \ {\itk A note on strongly additive conformal 
  field theory and half sided modular conormal standard inclusions}, 
  Lett.\ Math.\ Phys.\ {\bfk31} (1994) 303-308\sn
\q{65}E.\ Witten, {\itk Non-abelian bosonization in two dimensions}
   Commun.\ Math.\ Phys.\ {\bfk92} (1984) 455-472\sn
\q{66}A.B.\ Zamolodchikov, {\itk Integrable field theory {}from
    conformal field theory}, Adv.\ Studies in Pure Math.\ {\bfk19}
    (1989) 641-674\sn  
}  }\bye